\documentclass[12pt,notitlepage,a4paper]{article}
\usepackage{color,graphicx}
\usepackage{cite}
\usepackage{amssymb}
\usepackage{delarray,amsmath}
\usepackage[latin1]{inputenc}
\usepackage[american]{babel}
\usepackage[table]{xcolor}
\usepackage{cite}
\usepackage{hyperref}
\usepackage{url}
 \usepackage[bottom]{footmisc}

\newcommand{\be}{\begin{equation}}
\newcommand{\ee}{\end{equation}}
\newcommand{\beq}{\begin{equation}}
\newcommand{\eeq}{\end{equation}}
\newcommand{\beqa}{\begin{eqnarray}}
\newcommand{\eeqa}{\end{eqnarray}}

\newcommand{\bear}{\begin{eqnarray}}
\newcommand{\eear}{\end{eqnarray}}
\newcommand{\pd}{\partial}
\numberwithin{equation}{section}
\newcommand{\vev}{v.e.v.}

\newfont{\namefont}{cmr10}
\newfont{\addfont}{cmti7 scaled 1440}
\newfont{\boldmathfont}{cmbx10}
\newfont{\headfontb}{cmbx10 scaled 1728}

\begin{document}
\pagestyle{plain}
\setcounter{page}{1}

\begin{center}
\vspace{0.1in}

\renewcommand{\thefootnote}{\fnsymbol{footnote}}

\begin{center}
\Large \bf  Resonant Drivings in Global AdS
\end{center}
\vskip 0.1truein
\begin{center}
\bf{Javier Mas\footnote{javier.mas@usc.es} 
and  David Travieso Mayo\footnote{david.travieso.mayo@usc.es}\\
~}\\
\end{center}
\vspace{0.5mm}

\begin{center}\it{
Instituto Galego de F\'\i sica de Altas Enerx\'\i as (IGFAE) \\
Universidade de Santiago de Compostela  \\
E-15782 Santiago de Compostela, Spain}
\end{center}

\setcounter{footnote}{0}
\renewcommand{\thefootnote}{\arabic{footnote}}

\vspace{0.4in}

\begin{abstract}
\noindent
We revisit the case of a real scalar field in global AdS$_4$ subject to a periodic driving. We address the issue of adiabatic preparation and deformation of a time-periodic solution dual to a Floquet condensate. Then we carefully study the case of driving close to the normal mode resonant frequencies. We examine different slow protocols that adiabatically change the amplitude and/or the frequency of the driving. Traversing a normal mode frequency has very different results depending upon the sense of the frequency modulation. 
Generally, in the growing sense, the geometry reaches a periodically-modulated state, whereas in the opposite one, it collapses into a black hole. We study the suppression points. These are periodic solutions that are dual to a scalar field with vanishing $\vev, \langle \phi\rangle = 0$, instead of vanishing source. We also investigate quasi-periodic solutions that are prepared by driving with a combination of two normal resonant frequencies. We observe that, while the driving is on, the non-linear cascading towards higher frequencies is strongly suppressed. However, once the driving is switched off, the cascading takes over again, and in some cases, it eventually brings the solution to a collapse. Finally, we study the driving by a non-coherent thermal ensemble of resonant drivings that model stochastic noise. Our numerical results suggest the existence of stable regular solutions at sufficiently low temperature and a transition to collapse above some threshold.

\smallskip
\end{abstract}
\end{center}

\newpage

\tableofcontents

\newpage

\section{Introduction and main results}

The vast majority of the research performed in the context of the AdS/CFT correspondence is devoted to the study of static geometries which are dual to states of strongly coupled systems at equilibrium. The fact that holography extends, without further assumptions, into the realm of time-dependent situations cannot be overestimated. In this respect, many works have put their focus on the relaxation of initial perturbed geometries to static black holes. This should be dual to thermalisation on the field theory side. In \cite{Bizon:2011gg} a class of initial conditions were shown to end up collapsing and forming a black hole, even in the infinitesimal amplitude limit. With the study of \cite{Buchel_2013}, the situation was seen to be more involved, and the long-time behaviour of a perturbation started receiving much attention. 

In \cite{Carracedo2017,biasi2018floquet} we contributed to the program of simulating holographic dynamical open systems. Our setup involved the dual of a pure state in the form of a scalar field in global AdS. Other works addressed the driving of a holographic system in the deconfined phase \cite{Rangamani2015}. The main difference between these two setups is the balance of degrees of freedom. The deconfined case involves a dual bulk geometry with a horizon. The ingoing boundary conditions prevent the formation of a stationary state, and all the energy that is pumped by the driving into the system ends up heating it and increasing its entropy. This is in agreement with the O$(N^2)$ unbalance of degrees of freedom between the quantum theory and the driving bath, which makes the bulk act as a refrigerator that soaks all the energy that is pumped in. In contrast, in the confined phase, dual to global AdS, there are opportunities for long-lived driven solutions.

Of particular interest are the so-called periodically-driven or Floquet systems. The existence of stable {\em time-periodic solutions} (hereafter TPSs) in this context is of much interest as it stands out against the lore that predicts constant heating of the system. In \cite{biasi2018floquet} we mapped out the region of stability for both a complex and a real massless scalar in global AdS. For a sufficiently low amplitude of the driving, there exist TPSs at any frequency.

The present paper is structured as follows. In section \ref{gensetup} we briefly set up the stage. We review aspects of the previous work in \cite{biasi2018floquet} that will be important in the rest of the paper, particularly the notion of adiabatic preparation and transformation of a driven time-periodic solution (TPS) dual to a Floquet condensate. The next long section \ref{resondriv} contains the main results of our work. First, we will investigate the response of the system to drivings at resonant frequencies of the linearised system. Unlike the case of the linearised normal-modes, here the amplitude growth saturates and produces an interference pattern. When the driving is switched off, contrary to what happens out of resonance, where AdS is recovered, there generally remains a time-periodic oscillon solution without a source.

In later subsections we examine other protocols. For example, we chirp the frequency across a resonant value. We find an asymmetry in the oscillon that remains excited which is correlated with the sense of the modulation. In general, for a given (not too low) speed of the modulation, the remaining oscillon is more massive when the chirp is performed in the downward direction than the other way round. For sufficiently slow modulation speed, the oscillating solution collapses to a black hole geometry. 

The adiabatic preparation is a powerful method that we exploit in the rest of the paper in search of interesting stable dynamic solutions. This is the case of zero vacuum expectation value (hereafter $\vev$) solutions, which we try to prepare by an adiabatic modulation across frequencies where the linearised solutions exhibit this property. Unfortunately, the search for solutions is never perfectly prepared, and we discuss the possible reasons for that. Finally, we examine an exotic family of driven TPSs that have a mass (averaged over a period) lower than the one of AdS. This happens for driving frequencies below $\omega_\mathrm{b} = 1.25$.

We continue in section \ref{morefreq}, where we extend our method to drivings that involve two resonant frequencies in three scenarios, and we examine several protocols. In all cases, after switching off the signal, we find a quasi-periodic solution that is Fourier analysed for the spectral content. The Fourier analysis reveals an interesting property that did not show up in the single-mode case: the non-linearity excites a tower of higher modes, and not just the frequencies present in the soliton solution after the single-mode driving, which is something to be expected. What we find is that the cascading is blocked while the driving is active. In a sense, the open system boundary conditions act like a UV filter. Once the driving is switched off and the system goes back to isolation, the weak turbulence sets in, and a more intense cascading towards the UV is observed. In some cases, this cascading is so strong that the undriven solutions eventually tend to collapse in the long run, but we have not been able to follow the simulation up to this point. In a sense, this links to observed behaviour in Floquet systems where stability and locality in many body systems are more likely to be protected \cite{Sierant:2022xtl}.

In the last section \ref{multimode}, we press the organ's keyboard at full. That is, we introduce a multi-mode driving where the spectrum is that of thermal noise. In retrospect, we initiated the present paper motivated by the possibility of modelling the coupling of a pure quantum state to a thermal bath. The simulation is quite challenging, but we keep accurate values of the constraints under control as well as a high degree of convergence. The results extend the findings in the two-mode driving of the previous section. The driving keeps the cascading very much suppressed, and we indeed find solutions which show no signal of instability for a fairly long duration. This is by itself a remarkable result: the bulk fields, in particular the metric, evolve into a highly excited chaotic, yet pure, state. We can play with the temperature that controls the exponential damping of the noise spectrum. There seems to be a transition to an unstable solution above some value. This deserves further numerical study to elucidate the exact nature of the long-time behaviour.

We have deferred to Appendix A considerations about the reliability of the numerical implementation, including convergence tests. 

\section{General setup}
 \label{gensetup}

Our case study involves the simplest possible setup, namely a real massless scalar field in global AdS$_4$, 
\be
\label{com_act}
S=  \frac{1}{16 \pi G} \int d^{4}x \sqrt{-g}\left( R - 2\Lambda \right) - \frac{1}{2} \int d^{4}x \sqrt{-g} \pd_\mu\phi  \pd^\mu\phi \ ,
\ee
with  $\Lambda = -3/l^2$. From now on, we set $8 \pi G=2$. The ansatz for the metric is  
\be
ds^2 = \frac{l^2}{\cos^2 x}\left( - f e^{-2\delta} dt^2+ f^{-1} dx^2 + \sin^2 x \, d \Omega_{2}^2\right) \ , \label{line1}
\ee
where  $x\in [0,\pi/2)$ is the radial coordinate. This makes altogether a system of three functions $(\phi(t,x), f(t,x),\delta(t,x))$ to be solved for.\footnote{See Appendix D in \cite{biasi2018floquet} for further details.}

Expanding near the boundary $x=\pi/2$  and solving the equations of motion (see Appendix \ref{app: numerics}) yields:
\beqa
\phi(t,x) &=&\phi_{0}(t) -\frac{1}{2}\ddot \phi_{0}(t)( \pi/2-x)^2 +  \phi_{3}(t) ( \pi/2-x)^3 + {\cal O}(( \pi/2-x)^{4}) \ , \nonumber\\
\delta(t,x) &=&\frac{1}{2}\dot\phi^2_{0}(t)( \pi/2-x)^2 + {\cal O}(( \pi/2-x)^{4}) \ , \nonumber\\
f(t,x) &=& 1- \dot\phi^2_{0}(t)( \pi/2-x)^2+ f_{3}(t) ( \pi/2-x)^3  + {\cal O}(( \pi/2-x)^{4})\ ,  \label{expbound}
\eeqa
where we have chosen the proper time to be the coordinate time at the boundary. The coefficient function $f_3(t)$ is related to the instantaneous mass of the system and is given by a radial integral that is constructed with $\phi(t,x)$ and $\delta(t,x)$.
 The following constraint,
 \be
  \dot f_{3} = 6 \phi_{3}\dot\phi_{0} \ , \label{wardide}
 \ee
 is the $tt$ component of the conservation law for the energy-momentum tensor that follows from coordinate reparametrisation invariance. Finally, the function $\phi_3(t)$ is related by the holographic dictionary to the $\vev$ of the  operator ${ \Phi}$, dual to $\phi(t,x)$, and is fixed by demanding regularity of the full solution. 
 
The mere reference to resonance  motivates to start with the linearised theory in the background of global AdS. This is $f(t,x)=1, \delta(t,x)=0$ and $\phi(t,x)$ solving the linearised equation of motion. For a scalar field in AdS$_{d+1}$, the normal modes and their respective frecuencies are
 \begin{equation}
e^\pm_{n}(x) =\cos^{\Delta_\pm}x P_{n}^{(\frac{d}{2}-1, \Delta_\pm - \frac{d}{2})}(\cos(2x)) \ , ~~~~~~~\omega_{n} = 2n+\Delta_\pm \ ,  \label{linsol}
\end{equation}
with $\Delta_- =\Delta_+-d$ and $\Delta_+\Delta_-=0$. In our case, $d=3$ and $\Delta_- = 0$ or $\Delta_+=3$. Solutions $e_n^+(x)$ are normalisable and have vanishing source $e_n^+(\pi/2)=0$. Solutions $e_n^-(x)$ are non-normalisable and have vanishing $\vev$ $\left. \partial^d_x{e_n^{-}}(x)\right\vert_{x=\pi/2}=0$. 

\begin{figure}[ht!]
\begin{center}
 \includegraphics[scale=0.55]{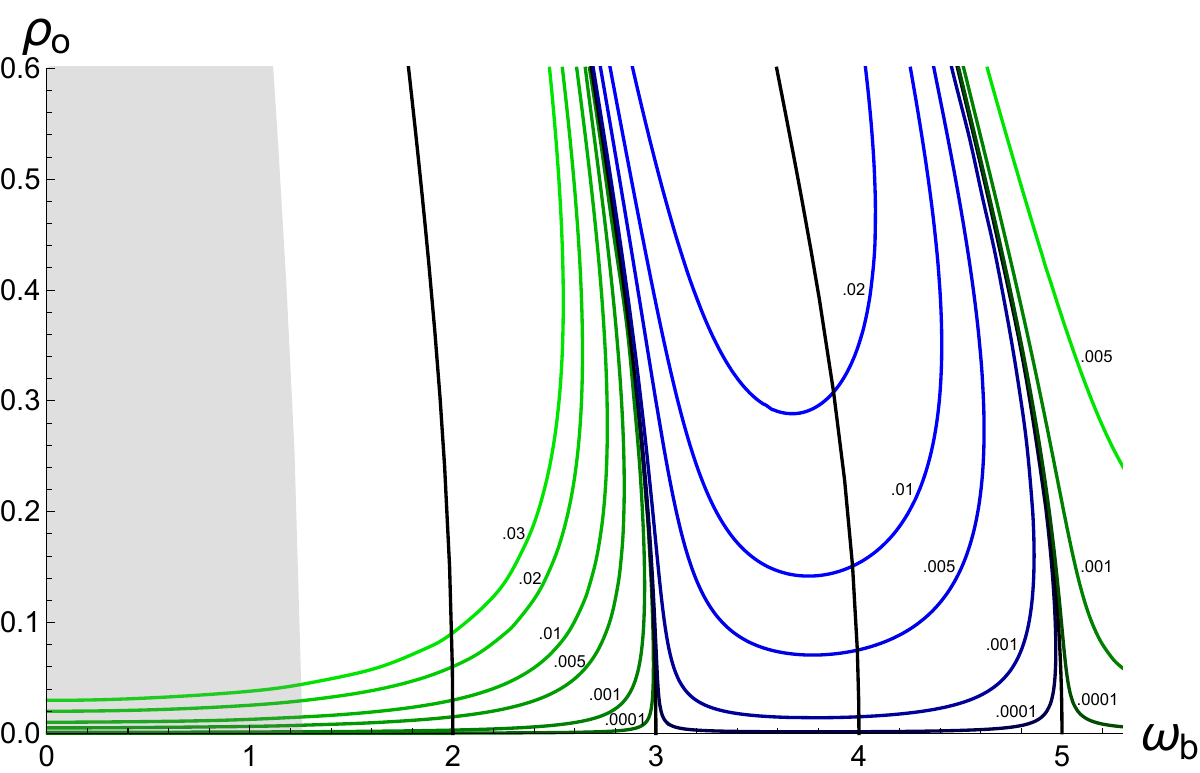}
\caption{Level plot showing the time-periodic solutions (TPSs). The horizontal axis is for the driving frequency $\omega_\mathrm{b}$ and the vertical axis is for the maximum amplitude at the origin $\rho_\mathrm{o}\equiv  \max_{t\in \mathrm{period}}\phi(t,0)$. The level curves are labelled by the value of the driving maximum amplitude  $\rho_\mathrm{b}$. The vertical black curves emerging from the frequencies $\omega_n = 2n+3$ correspond to oscillon solutions with vanishing driving amplitude, $\rho_\mathrm{b} = 0$. Those emerging from $\omega_n = 2n+2$ are TPSs with (almost) vanishing $\vev$ $\langle \phi\rangle \sim  0$ (see later in the text). The shaded region to the left signals the domain of TPSs whose average mass is negative $\bar M <0$. 
 }
\label{fig:level_plot}
\end{center}
\end{figure}

\begin{figure}
\begin{center}
\includegraphics[scale=0.6]{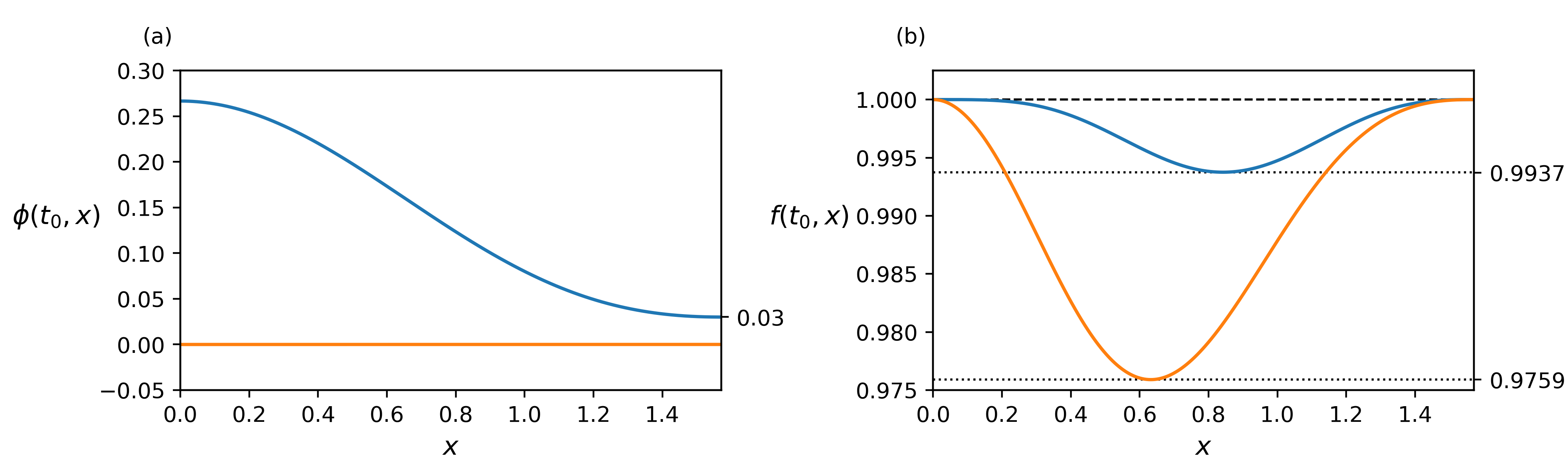}
\end{center}
\caption{Time-periodic solution (TPS) with  $\rho_\mathrm{b}=0.03$ and $\omega_\mathrm{b}=2.5$. The plots show two snapshots of the scalar field $\phi(t_0,x)$ (a) and the metric $f(t_0,x)$ (b) at two fixed times: one when the amplitude of the scalar field at the boundary reaches its maximum (blue) and another when it vanishes (orange). For this amplitude of the driving, the backreaction of the metric is very small, $f_\mathrm{min} = \min_{x\in(0,\pi/2)} f(t,x) = 0.9759$ and $1-f_\mathrm{min} =0.0241$. }
\label{fig:xxProfilesphif}
\end{figure}

The normalisable solutions $e_n^+(x)$ can be promoted to full non-linear oscillating solutions up to fairly high values of the amplitude \cite{Maliborski2013}. In this work we will term them {\em oscillons}. The pseudo-spectral construction in space is supplemented with a Fourier series decomposition in time to account for the proper periodicity. In \cite{biasi2018floquet} this construction was extended to non-normalisable, i.e. sourced, time-periodic solutions with arbitrary frequencies $\omega_\mathrm{b}\in {\mathbb R}$. The boundary condition needed is $\phi_\mathrm{b}(t) \equiv \phi(t,x=\pi/2) = \rho_\mathrm{b} \cos(\omega_\mathrm{b} t)$, where $\omega_\mathrm{b}$ and $\rho_\mathrm{b}$ are the {\em driving frequency} and {\em amplitude} respectively. One of the main results in \cite{biasi2018floquet} is the plot reproduced here in Fig. \ref{fig:level_plot}. Every point in the plot corresponds to a time-periodic solution (TPS) labelled according to $(\omega_\mathrm{b},\rho_\mathrm{o} = \max_{t\in \mathrm{period}}(\phi(t,0))$, namely, the driving boundary frequency and the maximum amplitude of the scalar field oscillation at the origin (see Fig. \ref{fig:xxProfilesphif}).\footnote{Notice that the oscillations $\phi(t,x)$ at any fixed value of $x$ are not harmonic, just periodic.} Level curves correspond to fixed values of the driving amplitude $\rho_\mathrm{b}$. The holographic dictionary relates these solutions to some time-dependent state of a periodically-driven Hamiltonian whereby $\phi(t,x=\pi/2)$ is the coupling.

The normalisable non-linear oscillons with vanishing driving, $\phi_\mathrm{b}=0$, span the black curved lines that hit vertically the horizontal axis at values of $\omega_\mathrm{b} = 2n+3$. On either side, each point corresponds to a TPS with one more node of the $\phi(x)$ profile on the right than on the left. Hence, they correspond to Floquet quantum phase transitions. On both sides, there are regions of stable solutions as well as regions of unstable solutions. The stability analysis was performed in \cite{biasi2018floquet}, where we refer the interested reader for further details.

\begin{figure}[h!]
\begin{center}
\includegraphics[scale=0.5, trim= 0 0.6cm 0 0, clip]{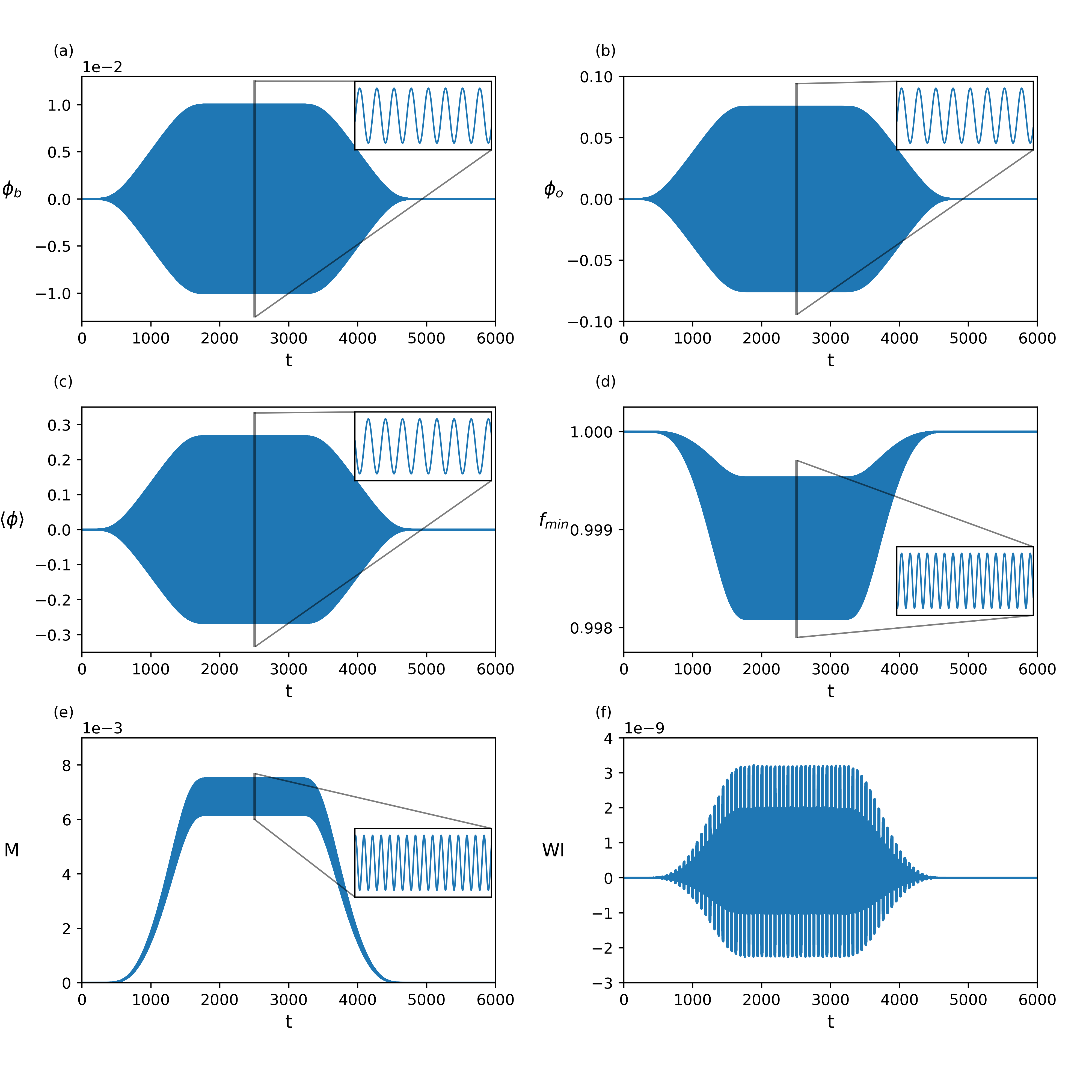}
\end{center}
\caption{Driving protocol for the creation of a time-periodic solution (TPS) with $\omega_\mathrm{b}=2.5$. (a) The driving amplitude rises adiabatically $\rho_\mathrm{b}: 0 \to 0.01$, and then smoothly turns off. (b) The scalar field at the origin $\phi_\mathrm{o}$, (c) the $\vev$ $\langle \phi\rangle$, (d) the minimum value of the metric function and (e) the mass oscillate periodically, signalling that we have indeed constructed a TPS. By slowly turning off the source, we return adiabatically to the vacuum state, where $\phi_\mathrm{o}=\langle \phi\rangle=M=0$ and $f_\mathrm{min}=1$. The metric, and therefore the mass, oscillates with double the frequency of the scalar field, to which it couples quadratically.}
\label{xQuench2d5plot}
\end{figure}

In that paper, besides the explicit numerical construction of the time-symmetric solution, another building method was shown to provide the same solutions by exploiting a holographic version of the adiabatic principle. Namely, starting from the AdS vacuum, the periodic driving is introduced by modulating the amplitude with a sigmoid-type function that slowly interpolates between zero and some finite value $(\omega_\mathrm{b},0)\to (\omega_\mathrm{b},\rho_\mathrm{b})$. The time span of this quench is labelled with $\beta$. For slow enough ramping, $\beta \gg 1$, the system follows a sequence of TPSs until the final one is reached. We will call such a protocol an {\em adiabatic injection}. The opposite one, where the driving is smoothly turned off at constant frequency, will be termed the {\em adiabatic extraction} protocol.

Not only the amplitude but also the frequency of the driving, if changed slowly enough, can interpolate smoothly between any two stable TPSs $(\omega_\mathrm{b},\rho_\mathrm{b})\to (\omega'_\mathrm{b},\rho'_\mathrm{b})$. This {\em frequency modulation} protocol allows for general paths to be traced in this parameter space. In this work, we want to see how things behave when these paths move close to or across critical lines of non-linear resonance (see Appendix \ref{app: numerics} for an explicit form of the modulation functions used in this paper). The TPS thus constructed is tantamount to a so-called {\em Floquet condensate} in the dual quantum field theory. The adiabatic preparation of Floquet condensates is a topic of active research in the context of periodically-driven systems \cite{Poletti2011,Heinisch2016,Weinberg2017}.

Fig. \ref{xQuench2d5plot} exhibits a collection of plots that will appear many times in this paper. Hence, we will pause here to guide the reader's eye. The first plot, (a), represents the driving profile. It starts from AdS and slowly builds up a harmonic driving with amplitude $\rho_\mathrm{b}=0.01$ and frequency $\omega_\mathrm{b}=2.5$. The solid blue colour is an artefact of the dense packing of the oscillations, as can be seen from magnifying the small rectangle. The time span of the injection envelope is $\beta = 2000$. 

In the remaining subplots, other relevant magnitudes have been monitored. Observe that they all oscillate periodically in time. Sometimes the period is halved. This happens for magnitudes that couple to $|\phi(t,\pi/2)|^2$, such as the metric function $f(t)$ and, hence, its minimum $f_\mathrm{min}(t)$ or the mass $M(t)$. As the magnified rectangles show, at intermediate times, we observe a fully developed TPS. The last plot, (f),  is a numerical check of the fulfilment of the Ward Identity, meaning the difference between the left-hand and right-hand sides in \eqref{wardide}. It is one of the consistency checks that we perform to ensure that the numerical solution is sound. This accuracy increases with the spatial grid resolution. In Appendix \ref{app: numerics}, we provide tests that exhibit fourth-order convergence, so we are very confident in the validity of our simulations. 

After the TPS has been built, the {\em adiabatic extraction protocol} smoothly follows the same sigmoid envelope but in reverse order. The end result is a graceful exit back into the vacuum AdS geometry. This smooth and reversible behaviour is generic as long as we keep away from the critical lines of oscillon solutions (see Fig. \ref{fig:level_plot}). There is where we expect, and indeed see, that the adiabaticity is lost, and a new phenomenon can be observed.

\section{Resonant driving}
\label{resondriv}

\begin{figure}[b!]
\begin{center}
\includegraphics[scale=0.55]{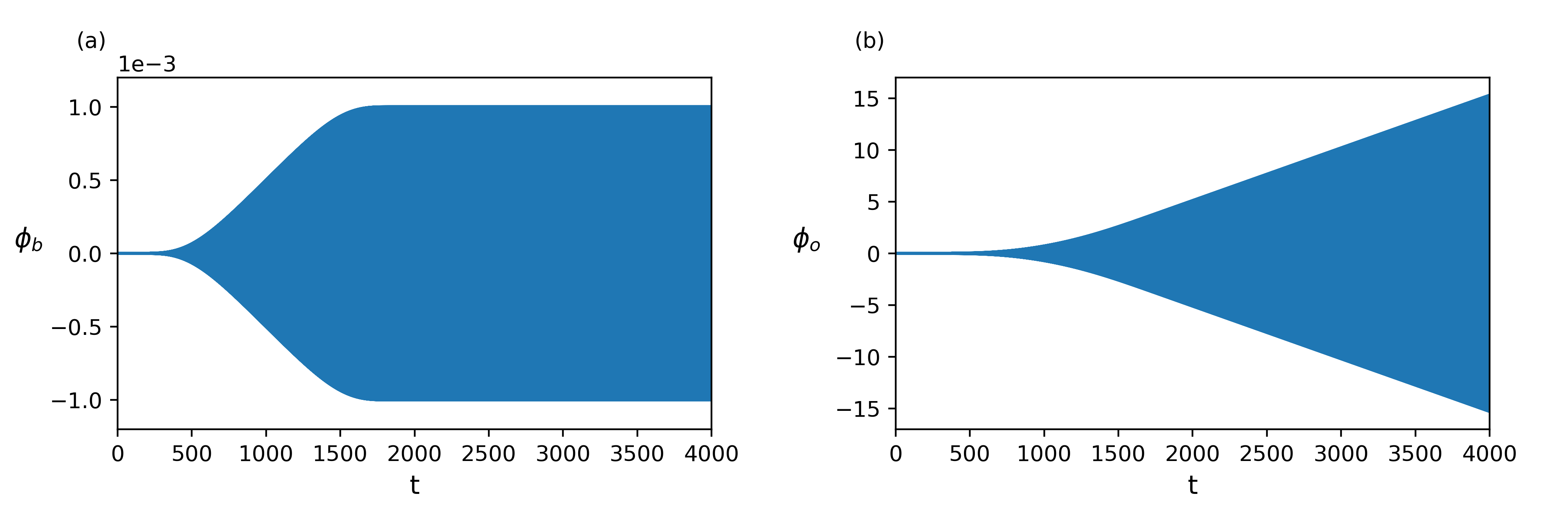}
\end{center}
\caption{(a) Driving at resonance $\omega_\mathrm{b}=3$ the scalar field on a background static AdS geometry. (b) The linear system experiences a standard resonance with a monotonic absorption of energy from the driving work.}
\label{xLinearQuenchPlot}
\end{figure}

In Fig. \ref{xQuench2d5plot} we have seen the response of the system to an adiabatic quench with a frequency far from resonance. Let us repeat this exercise at the first resonant frequency, namely $\omega_\mathrm{b}=3$. In order to make clear the importance of the backreaction, we will, first of all, drive the scalar field in a probe approximation on an AdS background. The equation is of course linear, and the result, as expected, leads to a monotonic increase in the amplitude. The result of the simulation is shown in Fig. \ref{xLinearQuenchPlot}. A much smaller injection, $\rho_\mathrm{b} = 0.001$, yields a linear growth of the oscillation amplitude with time $\rho_\mathrm{o}(t\gg 1) \sim t$.

\begin{figure}[hp!]
\begin{center}
\includegraphics[scale=0.5, trim= 0 1.5cm 0 1.5cm, clip]{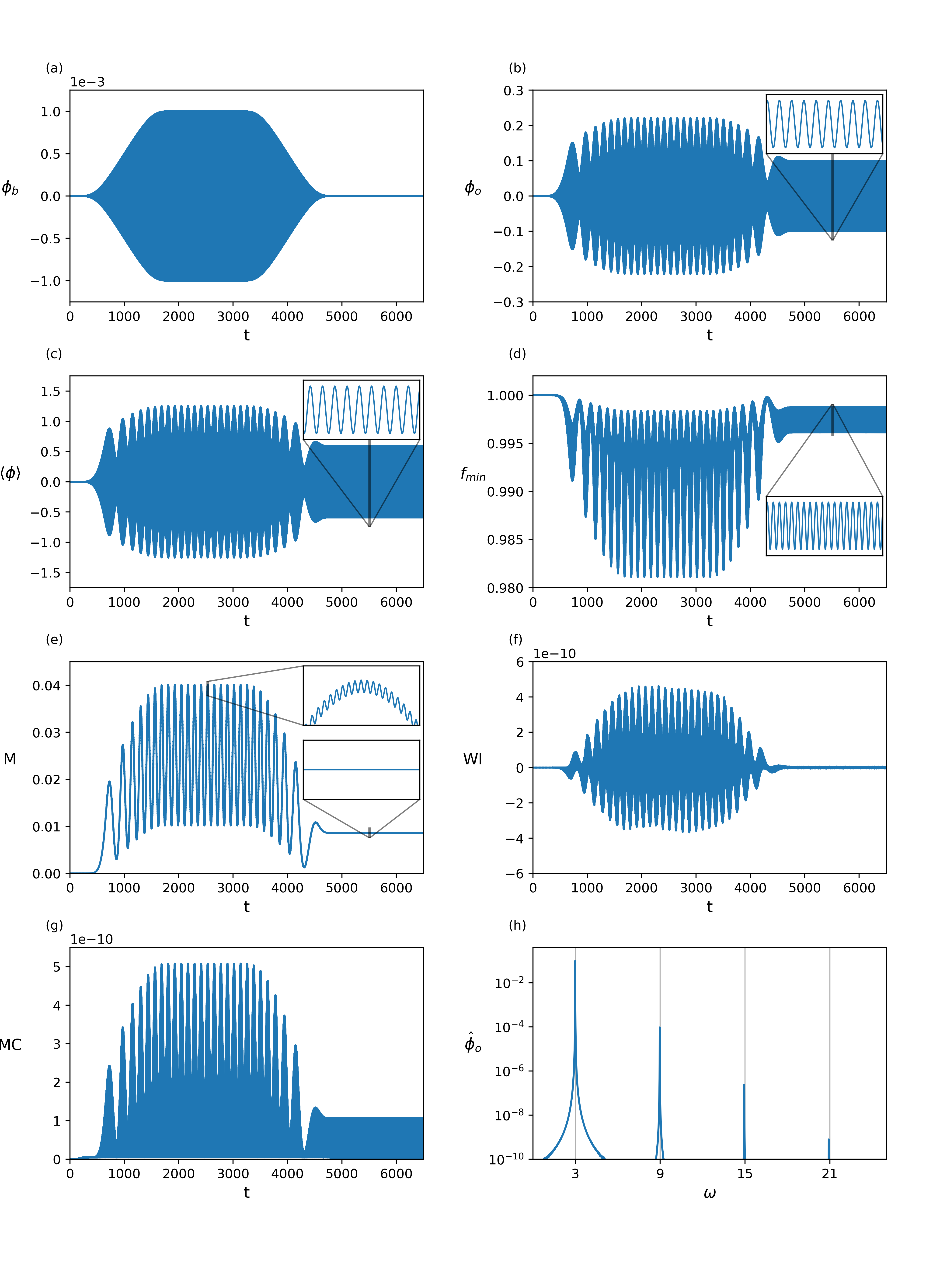}
\end{center}
\caption{(a) Driving at resonance $\omega_\mathrm{b}=3$, the scalar field excites a nearby frequency. (b-d) The interference pattern shows a strong amplitude modulation and the gain in mass saturates, (e). The accuracy of the simulation is backed by the smallness of the Ward Identity (f) and the integrated Momentum Constraint (g) (i.e. the spatially integrated absolute value of the difference between the left-hand and right-hand sides in \eqref{MC}). Whenever the driving is switched off, a sourceless TPS remains, namely, an oscillon. The Fourier analysis (h) reveals a tower of resonant frequencies, $\omega_\mathrm{osc}(2k+1)$ with $k\geq 0$, above the fundamental $\omega_\mathrm{osc} = 2.99$.  }
\label{xQuench3Plot}
\end{figure}

\begin{figure}[hb!]
\begin{center} 
\includegraphics[scale=0.6]{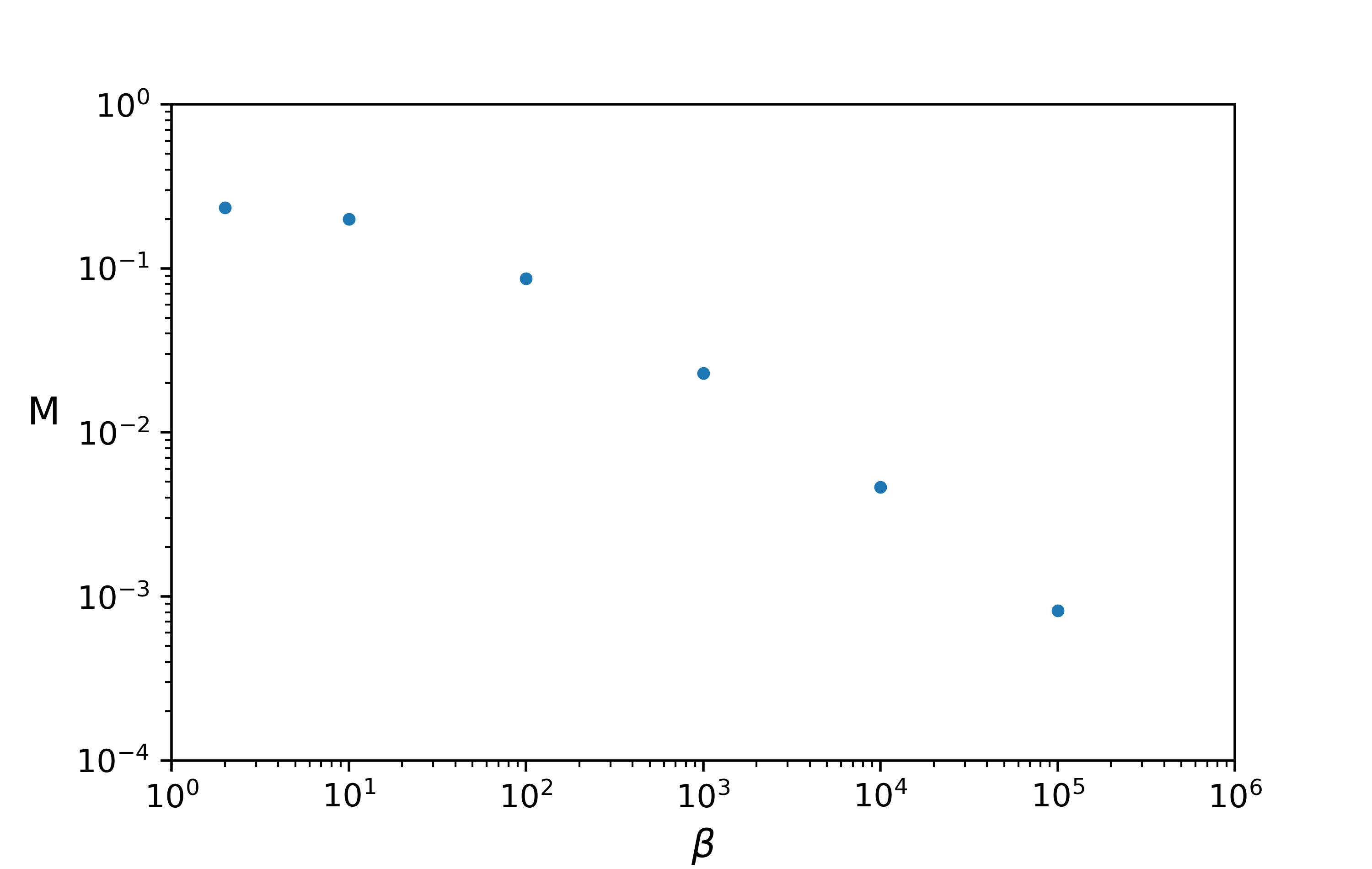}
\end{center}
\caption{Maximum mass of the remaining oscillon as a function of the adiabaticity parameter $\beta$. We see that in the limit $\beta \to \infty$, the end remnant mass vanishes, a signal that the geometry flows back to AdS, with a vanishing value of $\phi(t,x) = 0$. 
}
\label{max_fft_phio}
\end{figure}

The situation changes dramatically as soon as the gravitational coupling to the dynamical background geometry is switched on. This can be seen in Fig. \ref{xQuench3Plot}, with the same driving protocol as before. Now the response does not grow unbounded but saturates and develops a strong modulation in amplitude that follows the typical pattern of interference between two nearby frequencies. A Fourier analysis in time of the scalar field at the origin reveals what these two frequencies are. One is the driving frequency $\omega = \omega_\mathrm{b} = 3$. The other one is slightly lower and its shift is proportional to the amplitude of the driving. 

Around $t=3000$,\footnote{The natural time units are given by the driving period, $T=2\pi/ \omega \sim 2$ in this case. Hence, we are speaking roughly of over 1500 oscillations. } we turn off the driving again slowly and observe that the final solution {\em does not come back to AdS!} A non-linear oscillon with finite mass and vanishing source remains. 
In Fig. \ref{xQuench3Plot}(h) we have plotted the spectral content of the remaining oscillon. The Fourier analysis shows that the spectrum of modes is $\omega_k = \omega_\mathrm{osc}(2k + 1)$, where $k\geq 0$ and $\omega_\mathrm{osc} = 2.99$. It corresponds indeed to a periodic solution with period $T = 2\pi/\omega_\mathrm{osc}$. 

Which final remnant non-linear oscillon is left over varies with the particular protocol used for the driving process. In particular, it depends on the amplitude reached by the injection quench as well as on the speed and starting moment of the extraction phase. For a fixed maximum amplitude $\rho_\mathrm{b}$, its maximum possible final mass is smaller the slower we switch off the driving. In the infinitely slow limit, $\beta\to \infty$, the end state descends back to the static AdS geometry as in the case of a non-resonant driving. This can be seen in Fig. \ref{max_fft_phio}.

In a way, when we quench the system at a resonant frequency, the adiabaticity is lost, and the system does not return to the vacuum in finite time. In the spirit of the adiabatic theorem, this enhancement of the relaxation time should reveal a vanishing gap in the dual quantum theory. Indeed, we know the space of fluctuations of the non-linear oscillon geometries has one zero mode, precisely the one that moves along the line. 

\subsection{Frequency modulation across a resonance} 

\begin{figure}[hp!]
\begin{center}
\includegraphics[scale=0.5, trim= 0 1.5cm 0 1.5cm, clip]{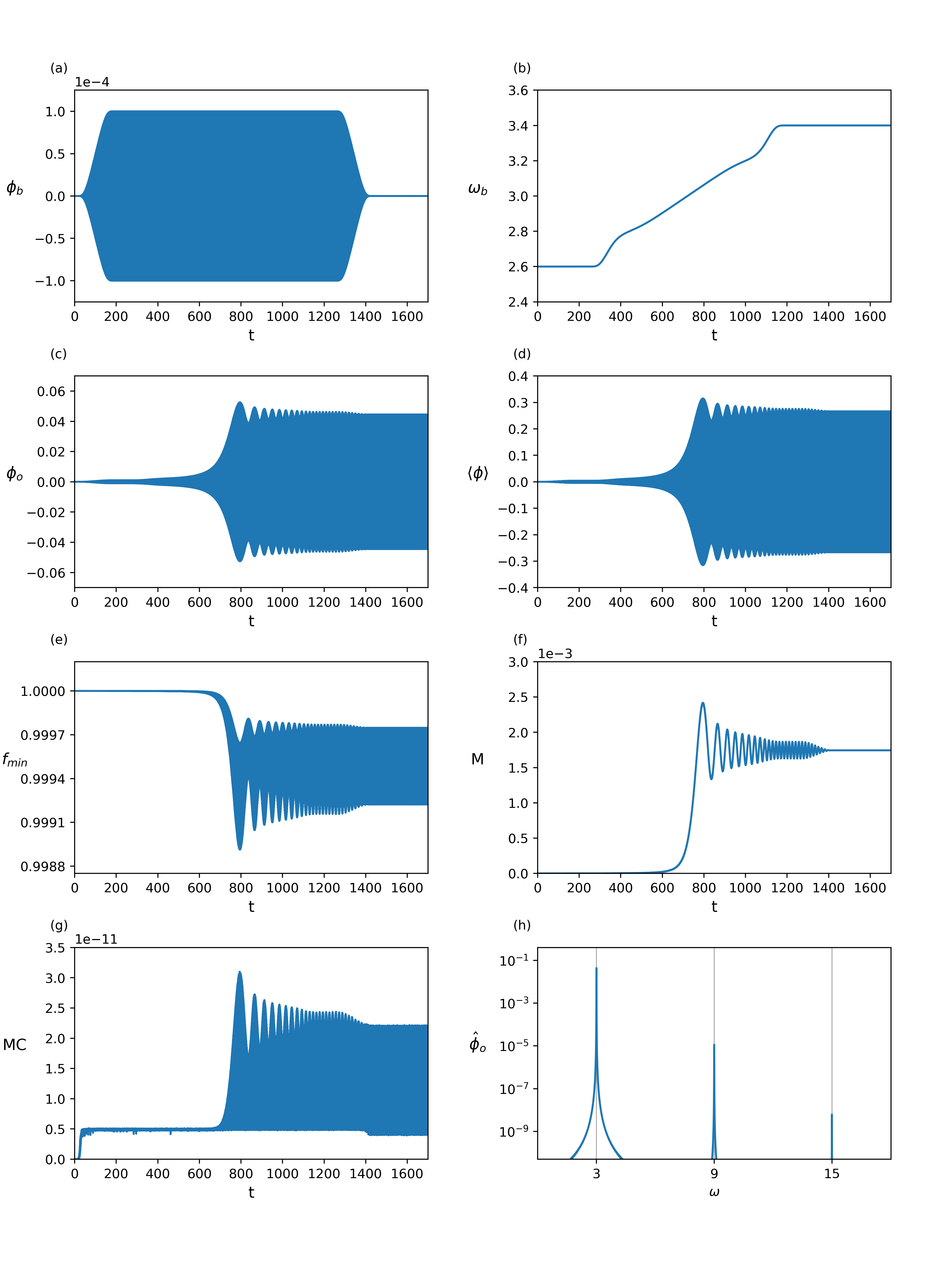}
\end{center}
\caption{Up-chirping. (a-b) A driving with a small value of the boundary amplitude $\rho_\mathrm{b} = 10^{-4}$ is injected at a frequency $\omega_\mathrm{b} = 2.6$. Then the frequency is modulated slowly ($\beta = 1000$) up to $\omega_\mathrm{b} =3.4$. Finally, the driving is extracted. (c-e) Close to $\omega_\mathrm{b} \sim 3$, the resonance excites a strong backreaction and a large gain in mass, (f). When the frequency of the driving is increased, the interference pattern with the excited oscillon becomes faster as their frequencies are split apart. After extraction of the driving, an oscillon remains, whose spectrum $\omega_\mathrm{osc}(2k+1)$, with $k\geq 0$ and $\omega_\mathrm{osc} = 2.998$, is shown in (h).
}
\label{xchirpup}
\end{figure}

\begin{figure}[hp!]
\begin{center}
\includegraphics[scale=0.5, trim= 0 1.5cm 0 1.5cm, clip]{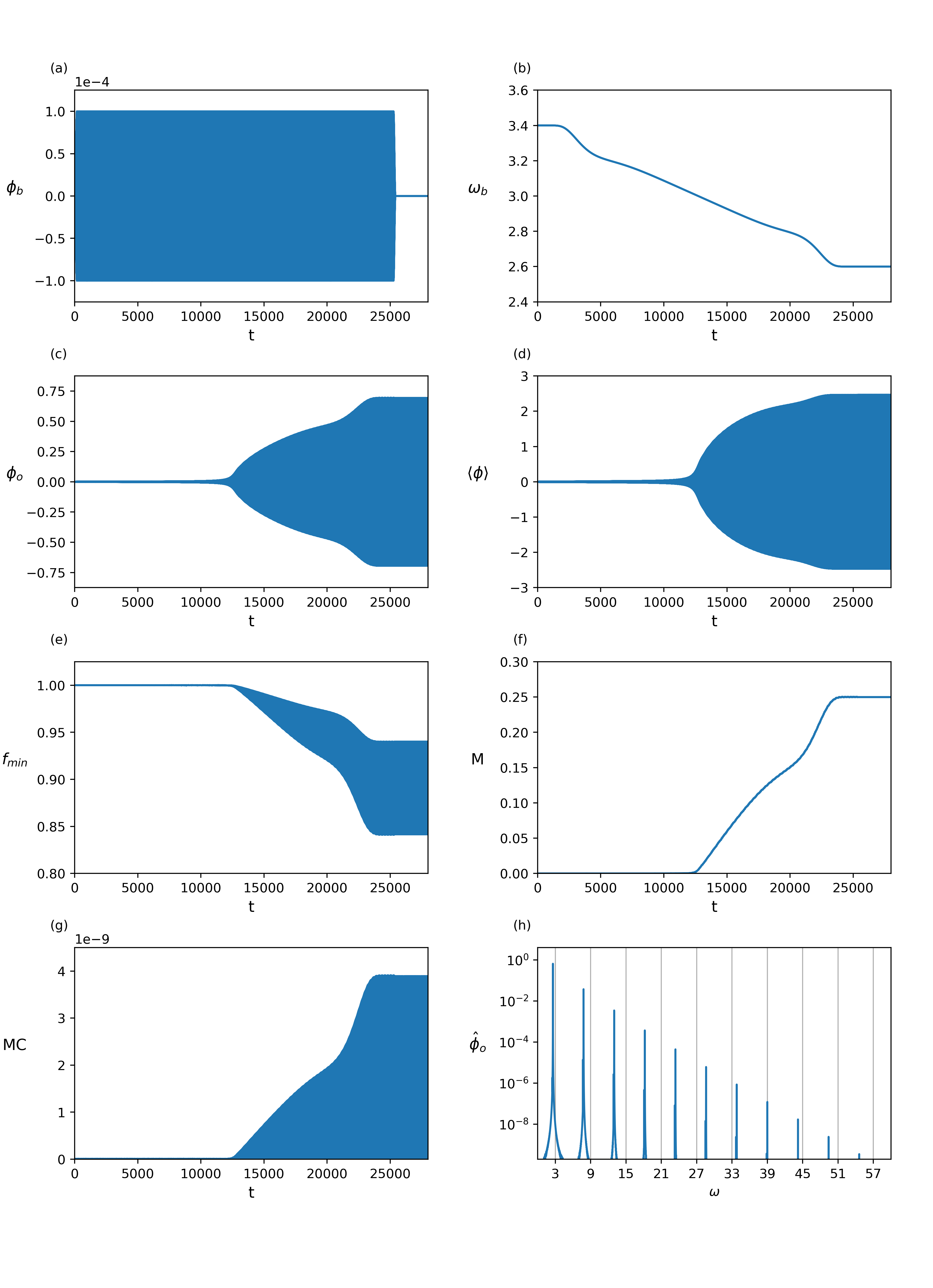}
\end{center}
\caption{Down-chirping. (a-b) A driving with a small value of the boundary amplitude $\rho_\mathrm{b} = 10^{-4}$ is injected at a frequency $\omega_\mathrm{b} = 3.4$. The frequency is modulated slowly downwards ($\beta = 2.5\times10^{4}$). As the linear resonant frequency $\omega_\mathrm{b} = 3$ is approached, the oscillation amplitude increases, as shown in  (c) and (d). This growth in amplitude also leads to an increment in the backreaction on the metric (e) and the mass (f).
We can stop the frequency down-shift whenever we want ($\omega_{\mathrm{b},\mathrm{final}} = 2.6$ in this case). After doing so, extraction of the driving (a) leaves an oscillon of, approximately, the end frequency of the chirp $ \omega_\mathrm{osc} \approx \omega_{\mathrm{b},\mathrm{final}}$. The thus-formed non-linear oscillon is much more massive than the one obtained in the up-chirping protocol. (h) The Fourier modes obey the spectrum $ \omega_\mathrm{osc}(2k + 1)$ with $k\geq 0$. 
}
\label{xchirpdownn}
\end{figure}

In this section, we shall explore the response of the system to a quasi-static modulation in frequency. The typical protocol will then involve three stages. Starting from AdS, we will first adiabatically {\em inject} a stable TPS of a given frequency $\omega_\mathrm{b}$. Next, the driving frequency will be {\em modulated} also very slowly by either increasing or decreasing it, according to the function given in \eqref{chirpfun}. After reaching some target value, the chirp will stop, and then an adiabatic {\em extraction} down to zero driving will be performed. Finally, we analyse the end state. The results can be succinctly condensed into the following statements:

- In the {\em growing} sense (Fig. \ref{xchirpup}), the state stops being a TPS while the driving frequency crosses the resonant value. After extraction, the solution settles down to an oscillon of some amplitude in the traversed branch. Which oscillon depends on the speed of the linearly growing modulation. In fact, this speed cannot be too slow; otherwise, the resonance will bring the TPS into the unstable region, and it will collapse. 

- In the {\em decreasing} sense (Fig. \ref{xchirpdownn}), for a slow modulation, the state follows a succession of TPSs that {\em never crosses} the oscillon line but asymptotically approaches it. The modulation can be stopped at a certain TPS that will lie close to the oscillon line. If the amplitude is now switched off, there will remain an oscillon of similar frequency and mass as the ones of the reached TPS. 

To clarify the last effect, notice by looking at Fig. \ref{fig:level_plot} that, while chirping down $\omega_{\mathrm{b},\mathrm{i}}\to \omega_{\mathrm{b},\mathrm{f}}$, the state moves adiabatically along a succession of TPSs, following a line of constant $\rho_\mathrm{b}$ that makes the amplitude at the origin, $\rho_\mathrm{o}$, grow. In this flow, these iso-curves of constant $\rho_\mathrm{b}$ to the right of an oscillon line asymptote to it. If the down-chirp stops at some frequency, switching off the driving amplitude makes the state land on the resonance line at an oscillon with approximately the frequency $\omega_{\mathrm{b},\mathrm{f}}$. The precision of the frequency reached in this way can be improved by following a lower $\rho_\mathrm{b}$ curve, which lies closer to the oscillon line.

\begin{figure}[t!]
\begin{center}
\includegraphics[scale=0.55]{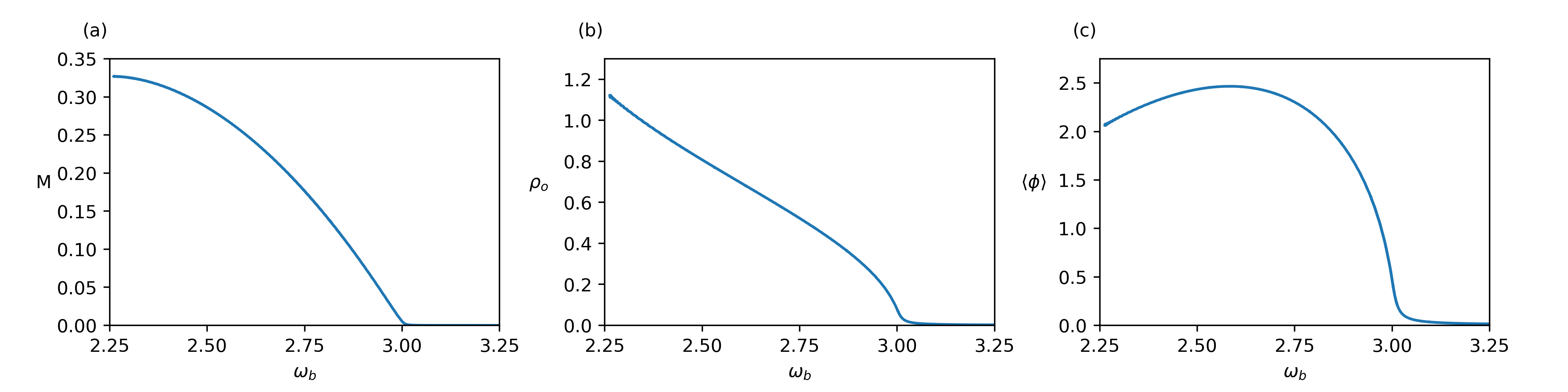}
\end{center}
\caption{Chirping down until collapse. The process starts at $\omega_\mathrm{b} = 3.25$ with $\rho_\mathrm{b} = 10^{-4}$ and chirps the frequency down without stopping. (a) As soon as the mass curve reaches a maximum, the system becomes unstable and collapses into a black hole. (b) As the frequency is modulated $\rho_\mathrm{o}$ builds the oscillon curve. (c) We also show the $\vev$ evaluated on this curve.}
\label{xoscillon_curve}
\end{figure}

Another very nice benefit of this observation is that, with this method, we can trace the complete oscillon curve. To achieve this, we simply need to keep moving down in frequency at a constant $\rho_\mathrm{b}$. For example, in Fig. \ref{fig:level_plot}, the blue level curve labelled with $\rho_\mathrm{b} = 0.0001$ becomes indistinguishable from the oscillon curve beyond $\rho_\mathrm{o}= 0.05$. In Fig. \ref{xoscillon_curve} we plot the full level curve and two observables evaluated on it: the mass $M$ and the $\vev$ $\langle \phi \rangle$. In (a) we observe the bending of the mass curve, $M(\omega_\mathrm{b})$, which eventually stops precisely when its slope becomes zero. This end point is nothing but the {\em Chandrasekhar} limit for oscillon instability 
towards black hole formation. The curve $\rho_\mathrm{o}(\omega_\mathrm{b})$ shown in (b) is the full development of the oscillon curve, of which in Fig. \ref{fig:level_plot} we only see the bottom part branching out from $\omega = 3$. The convex bending of the curve in Fig. \ref{xoscillon_curve}(b) around $\omega = 2.4$ could not be guessed from that plot. 

\begin{figure}[hb!]
\begin{center}
\includegraphics[scale=0.6, trim= 0 0.04cm 0 0.04cm, clip]{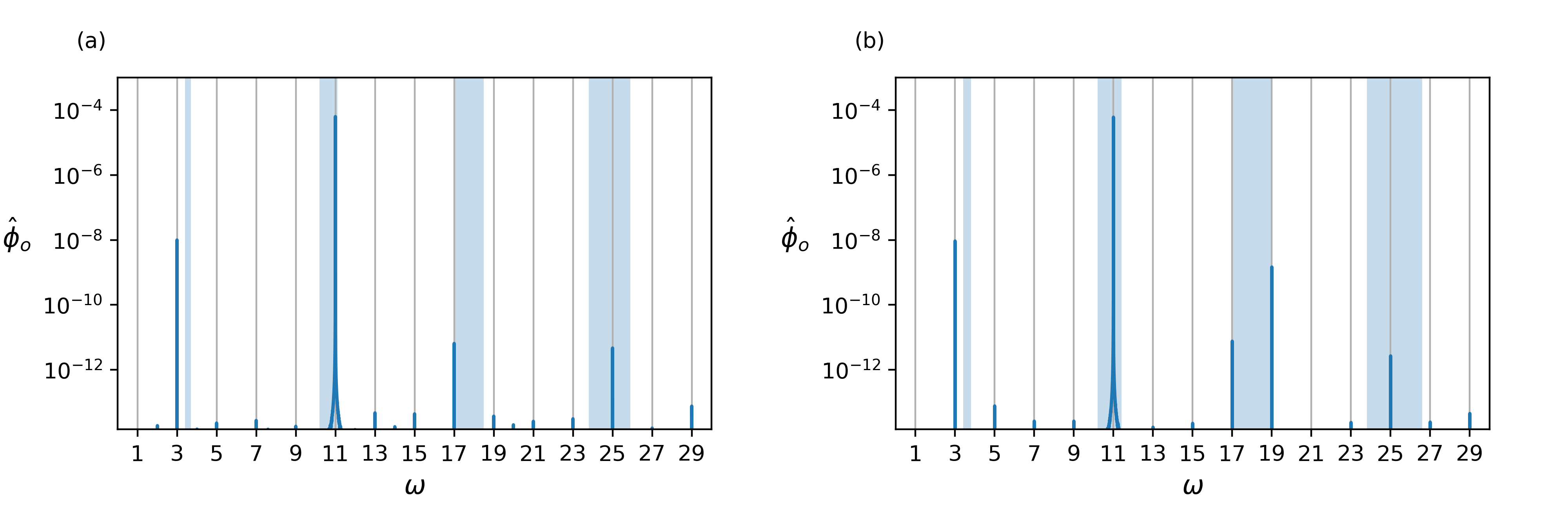}
\end{center}
\caption{In these plots, the driving frequency was modulated downwards in the interval $\omega_\mathrm{b} = 3.7\to 3.4$ (a) and $3.8\to3.4$ (b) with a driving amplitude of $\rho_\mathrm{b}=0.01$. The light blue columns show the intervals that have been swept by the fundamental and higher harmonics of the sourced modulated TPS. The vertical segments are the amplitudes of the Fourier modes in the remaining sourceless scalar field state after switching off the driving. We explicitly see that the Fourier mode at $\omega = 19$ was not excited in (a), but it was excited in (b). Hence, the final spectrum of linearised modes carries memory of the protocol performed in the past.}
\label{chirp_no_line}
\end{figure}

For the sake of completeness, we also mention here the case when the modulation in frequency does not traverse any oscillon line. When the process is performed without crossing any resonance curve, the system passes through a series of TPSs adiabatically. It seems plausible to expect a return to AdS when the driving is slowly turned off. This was the case without chirping. The situation is however richer now due to the fact that, remember, TPSs are just periodic solutions. Higher Fourier modes will also drift following the instantaneous spectrum $\omega_k(t) = \omega_\mathrm{b}(t)(2k+1)$. Very remarkably, if $\omega_\mathrm{b}(t)$ spans an interval of width $\Delta_0 = \omega_\mathrm{b}(t_\mathrm{f})-\omega_\mathrm{b}(t_\mathrm{i})$, the interval covered by the higher resonances will be wider, $\Delta_k = \Delta_0(2k+1)$. Most likely, many of these higher harmonic frequencies will drift across higher resonance lines, thereby exciting oscillons at that frequency. We can see this effect in Fig.  \ref{chirp_no_line}. Upon switching off the driving, the geometry indeed flows back to AdS, but the scalar field state will be populated with a linear combination, a `gas', of linearised normal modes that have been excited but do not backreact because of their smallness. The amplitudes of the remnant linearised modes carry a late-time memory of the chirping process that was performed.

\subsection{Suppression points: TPSs with vanishing $\vev$ }

As mentioned before, at the linearised level, the equations of motion admit two types of solutions (see \eqref{linsol}). Those named $e^+_n(x)$ have vanishing source and frequency $\omega_n = 2n+3= 3,5,...$ for AdS$_4$. On the other hand, we can find $e^-_n(x)$ with vanishing $\vev$, $\left. \partial^{(3)}_x{e_n^{-}}(x)\right\vert_{x=\pi/2}=0$, and frequencies $\omega_n = 2n+2= 2,4,...$

Most of the discussion so far has involved the response and construction of the non-linear solutions branching from the first type of normal modes $e_n^+$. 
In this section, we would like to address the second type of normal modes $e_n^-$. The line of non-linear oscillons with vanishing $\vev$ that branches out from the even frequencies can be found by an adiabatic modulation of the frequency, which eventually crosses this line. This is the same protocol that we used to cross the non-linear oscillon type. Here, however, the target is to find the values of $\rho_\mathrm{b}$ and $\omega_\mathrm{b}$ that produce a solution with $\langle \phi\rangle = 0$. We will name such solutions {\em suppression points} \cite{Berenguer_2022}.

\begin{figure}[ht!]
\begin{center}
\includegraphics[scale=0.5, trim= 0 0.6cm 0 0.35cm, clip]{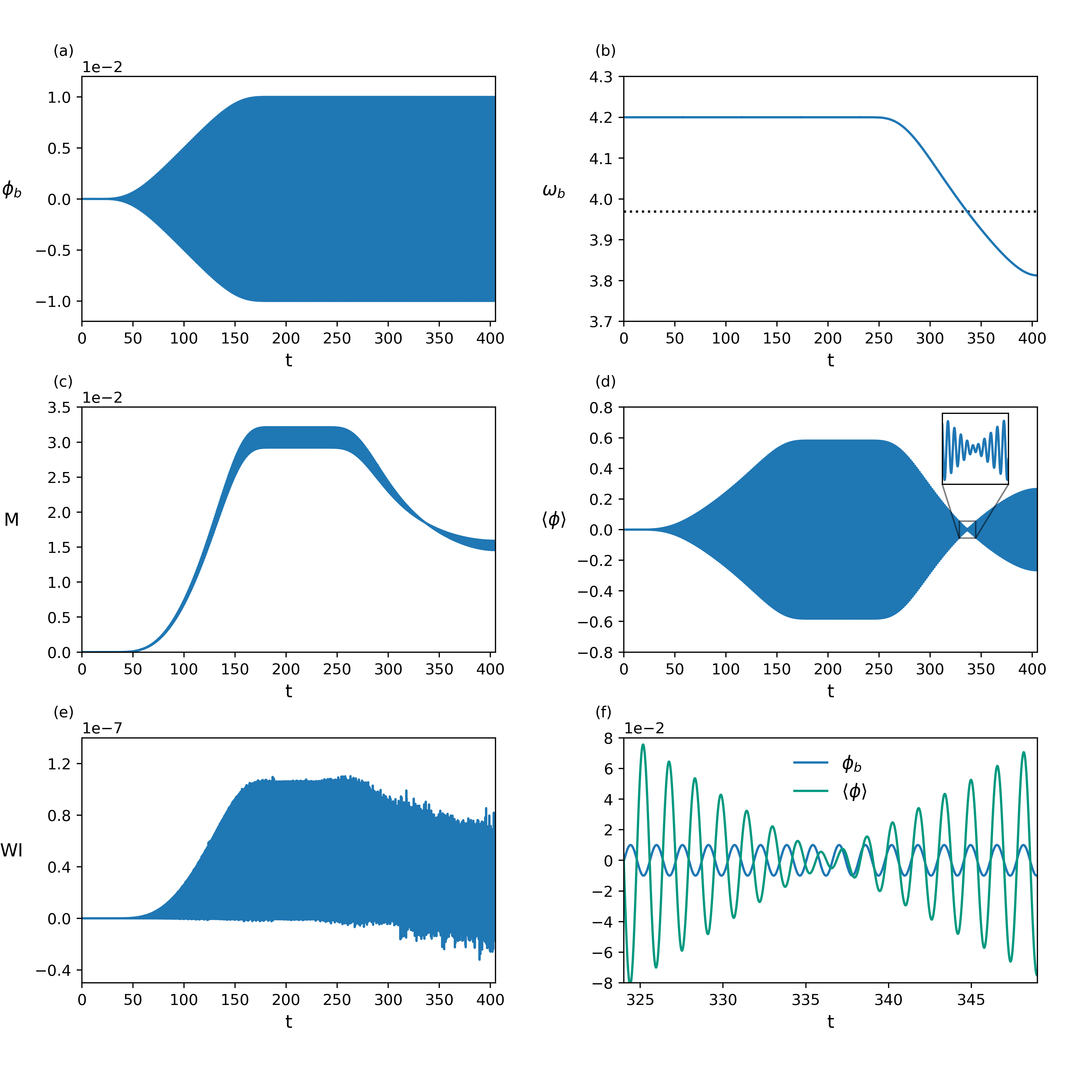}
\end{center}
\caption{Construction of the zero $\vev$ solutions. In the upper plots (a) and (b), we see the control variables. While $\rho_\mathrm{b}$ stays constant after a ramping time of $\beta=200$, the frequency $\omega_\mathrm{b}$ is modulated downwards, traversing a frequency (dotted line) where the amplitude of the oscillations in the mass (c) and in the $\vev$ (d) vanish. The inset in (d) enhances the view of the response $\vev$ $\langle \phi\rangle$, and we can see that the suppression is not exact. Plot (f) shows that there is a shift of $\pi$ in the relative phase between $\phi_\mathrm{b}$ and $\langle \phi\rangle$ from one side of the (almost) vanishing point to the other.}
\label{xzero_vev}
\end{figure}

The protocol is illustrated in Fig. \ref{xzero_vev} and explained in the caption. Indeed, the $\vev$ $\langle \phi\rangle$ appears to cross a vanishing point. The resulting locus spans a curve that can be seen in the level plots in Fig. \ref{fig:level_plot}, branching out vertically from even frequencies $\omega_\mathrm{b} = 2,4,...$ A closer look at that point shows that this is only nearly true. The amplitude of the oscillations in $\vev$ $\langle \phi(t)\rangle$ never actually vanishes. Interestingly, they instead exhibit a transition where there is a shift of $\pi$ radians in their relative phase with respect to the driving $\phi_\mathrm{b}(t)$ when passing from one side of the minimum to the other.\footnote{This phase shift is the dual of the analogous one that happens between the boundary driving $\phi_\mathrm{b}(t)$ and the oscillation at the origin $\phi_\mathrm{o}(t)$ when traversing the oscillon lines. In this case, this signals the presence of an additional node in the instantaneous profile function $\phi(t,x)$. } The cleanest protocol to construct the mentioned curves in Fig. \ref{fig:level_plot} is to perform small modulations in amplitude and frequency following the direction in which the $\vev$ is smaller, trying to move along that curve.

So there seems to be an asymmetry here between the oscillons (vanishing $\phi_\mathrm{b}(t) = 0$) and these suppression points (almost vanishing $\vev, \langle \phi\rangle$). Our interpretation of this result comes from examining the oscillon solutions. There, as said, the source vanishes, but the oscillation $\phi(t,x)$ and, in particular, the $\vev$ $\langle \phi(t)\rangle \neq 0$ are periodic yet non-harmonic functions. It is natural to expect what the dual of this statement should be: solutions with vanishing $\vev$, $\langle \phi(t)\rangle = 0$, should have a periodic but {\em non-harmonic source}, $\phi_\mathrm{b}(t) = \phi_\mathrm{b}(t+T)$, whose spectrum then involves Fourier modes $\omega_\mathrm{dual \, osc} (2k+1)$. Here $\omega_\mathrm{dual \, osc} \lesssim \omega_n = 2n + 2$ follows one of the black curves above the even frequencies in Fig. \ref{fig:level_plot}. The exact construction of these solutions remains open, but it should be feasible using the methods used in \cite{Maliborski2013,Maliborski2016}.

\subsection{Extracting energy out of AdS}

\begin{figure}[ht!]
\begin{center}
\includegraphics[scale=0.6]{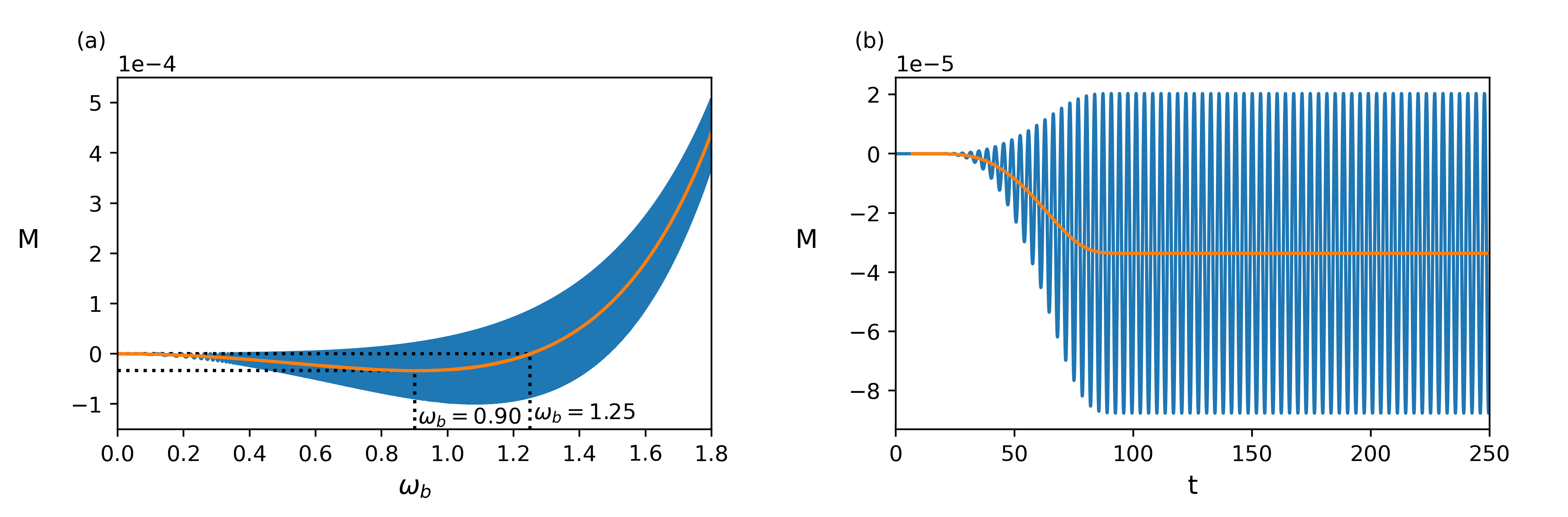}
\end{center}
\caption{Energy-extracting solutions. We plot the mass oscillation range in blue and the average mass in orange. (a) We have scanned over frequencies in the range $\omega_\mathrm{b} \in [0, 1.8)$ with a driving amplitude of $\rho_\mathrm{b}=0.01$, and we find that the TPSs with $\omega_\mathrm{b} \leq 1.25$ all have an average mass below zero. (b) We plot the mass for the TPS with $\omega_\mathrm{b}=0.9$ and $\rho_\mathrm{b}=0.01$ obtained after an {\em adiabatic injection} at a constant frequency.} 
\label{fig:xMasa_negativa}
\end{figure}

So far, all the TPSs have masses that oscillate in mean above the AdS mass (see, for example, Fig. \ref{xzero_vev}.) We say that they are {\em energy injecting}. This agrees with the idea that AdS is the geometry that corresponds to the ground state in the dual theory, whereas, for example, oscillons are dual to some kind of excited state in the boundary QFT. 
For TPSs, we do not expect a notion of dual state to be correct as these are sourced geometries. Still, the notion of average mass makes sense by continuity. For example, the sourced solutions that approach oscillons smoothly end up in such states upon switching off the source, where the average mass approaches the final constant mass of the oscillon.

\begin{figure}[h!]
\begin{center}
\includegraphics[scale=0.7, trim= 0 0.5cm 0 0.5cm, clip]{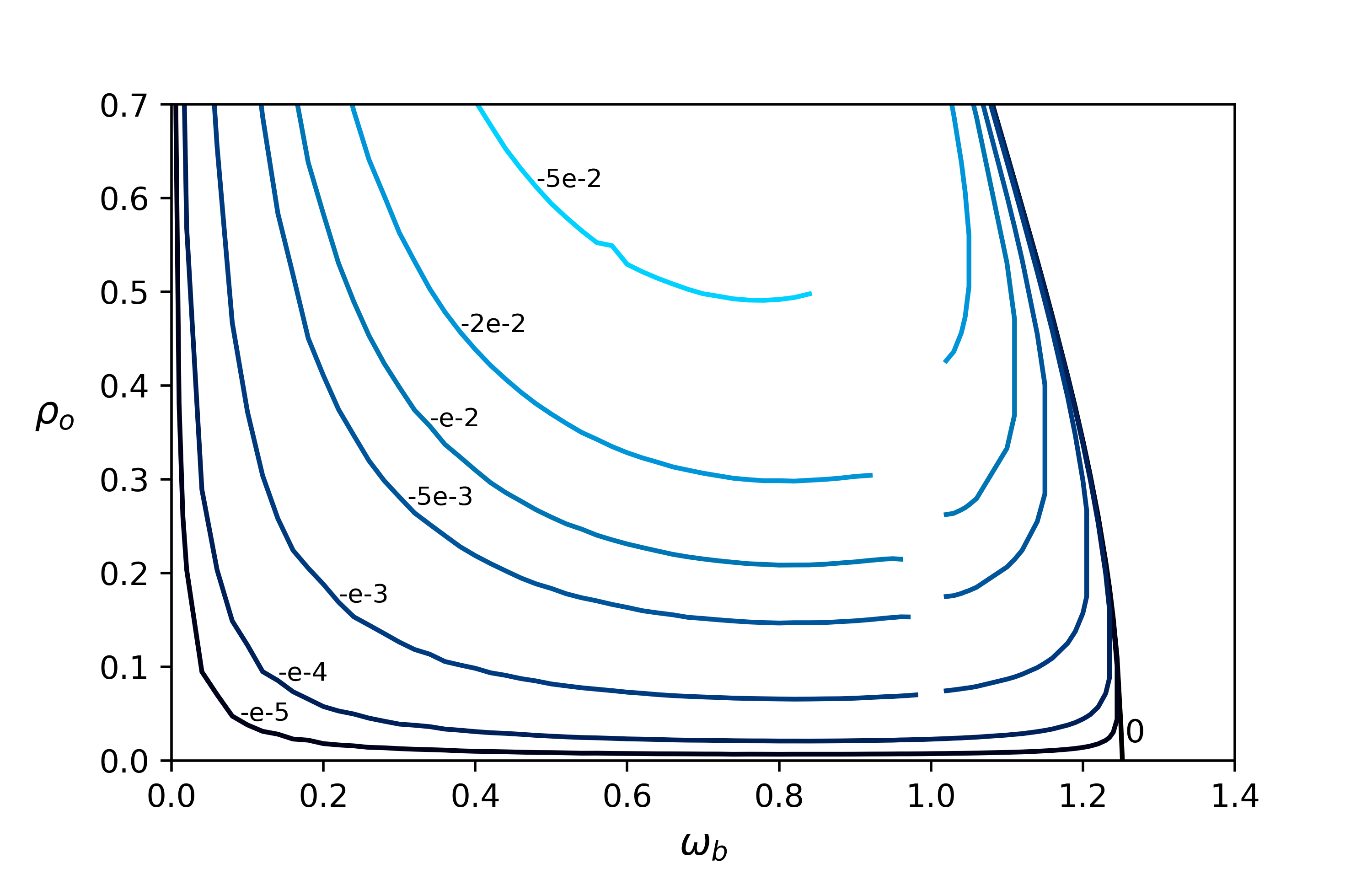}
\end{center}
\caption{Level curves  of (constant negative) average mass in the $(\omega_\mathrm{b},\rho_\mathrm{o})$ plane  inside the shaded region in Fig. \ref{fig:level_plot}. Each curve is labelled by the value of the mass. To construct a pumping solution, it is necessary to take a double scaling limit where $\omega_\mathrm{b}\to 0$ and $\rho_\mathrm{b}\to \infty $. In contrast, reaching the left edge of this plot ($\omega=0$) with a finite amplitude $\rho_\mathrm{o}$ makes the solution go back to $M=0$, as suggested by the level curves.}
\label{fig:xcurvas_Masa_negativa}
\end{figure}

In Fig. \ref{fig:xMasa_negativa}(a) we show the collection of TPSs that one can find in the interval $\omega_\mathrm{b} \in[0,1.8)$ with a driving amplitude of $\rho_\mathrm{b}=0.01$. The blue band is the range of oscillation of the mass $M$ at each TPS. We plot the average mass in orange and observe that its value is negative for $\omega_\mathrm{b}\leq 1.25$. Another way of constructing these solutions is by performing an {\em adiabatic injection} at a constant frequency (the same protocol used in section \ref{resondriv}). In plot (b) we show the mass for the TPS with $\omega_\mathrm{b}=0.9$ and $\rho_\mathrm{b}=0.01$ obtained by this method. In Fig. \ref{fig:xcurvas_Masa_negativa} we see this region of TPSs with negative average mass (shaded region in Fig. \ref{fig:level_plot}) and the level curves of constant mass. Unfortunately, there is no oscillon state near these frequencies, where we could gracefully switch off the driving, leaving a negative mass periodically oscillating solution. If there were such a sourceless negative mass oscillon, we would have indeed found a bona fide `holographic time crystal'.

This energy extraction behaviour can be continuously connected to the {\em pumping solution} obtained in \cite{Carracedo2017}. The pumping solution is a solution in which the source of the massless scalar field increases linearly in time at a constant speed, $\phi_\mathrm{b}(t) = \alpha_\mathrm{b} t$. It can be obtained from the TPSs in a double scaling limit: $\omega_\mathrm{b}\to 0$ and $\rho_\mathrm{b}\to \infty $ with $\omega_\mathrm{b}\rho_\mathrm{b} = \alpha_\mathrm{b}$ (see Appendix B in \cite{biasi2018floquet}). Pumping solutions can acquire a negative mass as long as the source $\alpha_\mathrm{b} \neq 0$ is maintained. This negative value flows back to zero or a positive value as soon as it is switched off. It is relevant at this point to mention the work in \cite{Ishii_2022}, where periodic driving is applied to a charged black hole dual to a thermal (positive mass) state. In that scenario, it is possible to have a net mass-energy extraction after switching off the periodic driving.

\begin{figure}[ht!]
\begin{center}
\includegraphics[scale=0.5, trim= 0 0.6cm 0 0.35cm, clip]{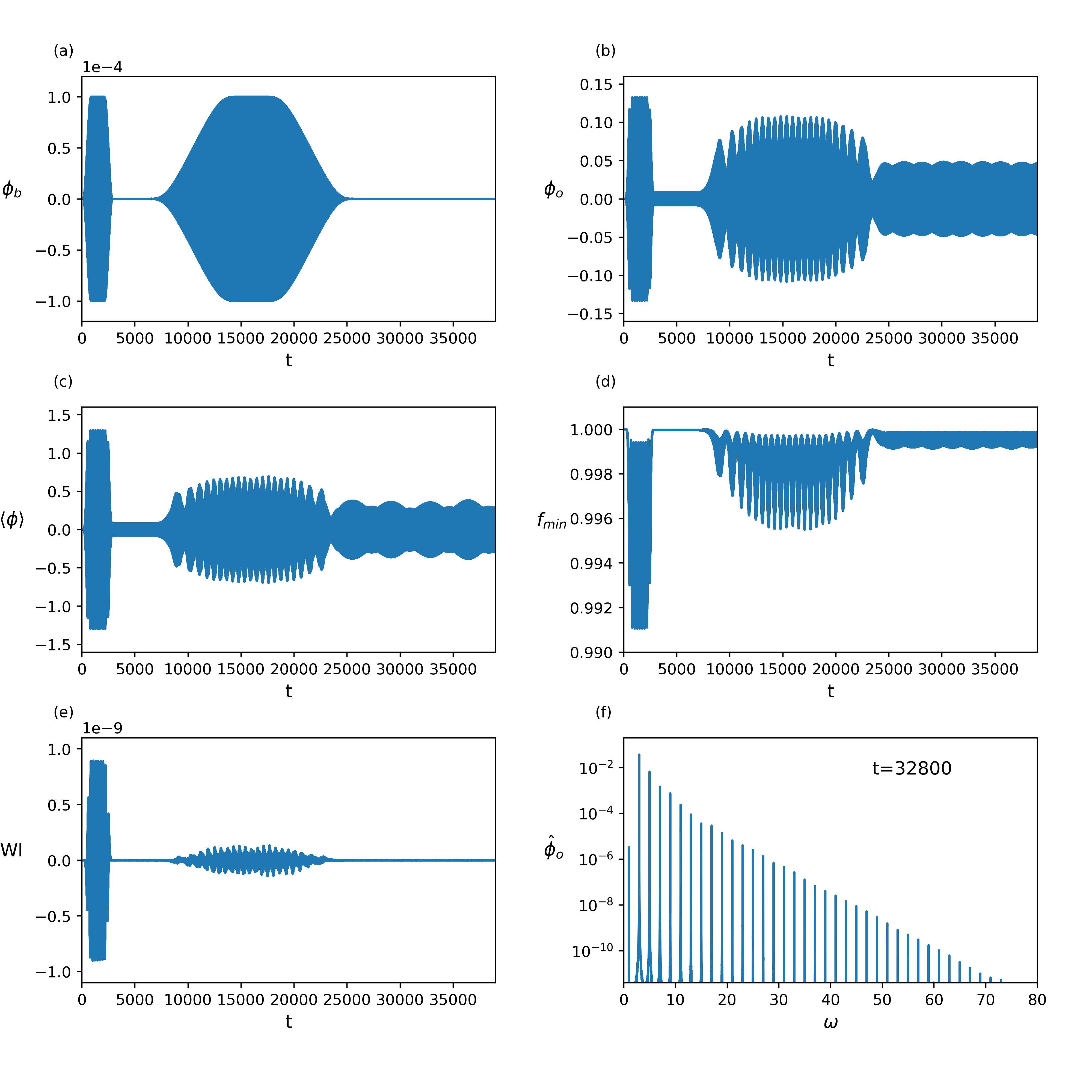}
\end{center}
\caption{(a) We create an $\omega=5$ oscillon and, on it, we inject and extract an $\omega=3$ periodic perturbation. (b-d) An interference pattern is produced that, after fully extracting the driving, relaxes to the remnant oscillation, showing a quasi-periodic profile for $t\geq 2.5\times10^{4}$. (e) The Ward Identity is always well satisfied, even with a rather sparse grid of $2^{10}$ points. (f) The time Fourier transform shows an exponential fall-off in the spectrum, where the frequencies are now the frequencies $\omega_k =2k+1 = 1,3,5,...$ The central time is indicated, and the time window for the Fourier transform is $\Delta{t}=10^4$.
}
\label{xQuench3and5Plot}
\end{figure}

\begin{figure}[ht!]
\begin{center}
\includegraphics[scale=0.5, trim= 0 0.6cm 0 0.35cm, clip]{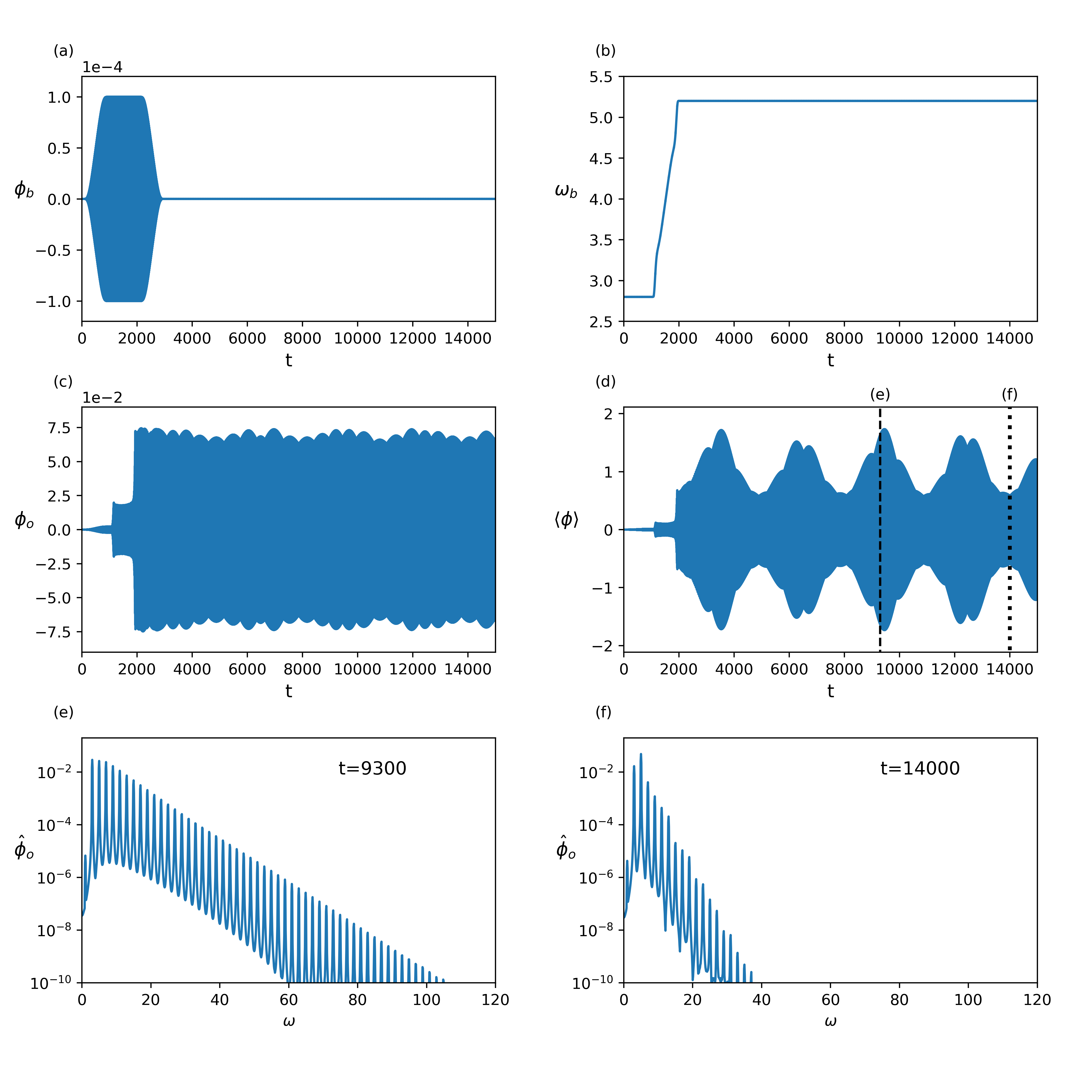}
\end{center}
\caption{The driving protocol shown in plots (a) and (b) involves a chirp in frequency at fixed amplitude from $\omega =2.8$ up to $\omega = 5.2$ during a time span of $\beta=1000$. Afterwards, the driving is extracted, and the signal is analysed. (c-d) The undriven solution evolves with a quasi-periodically modulated amplitude. (e-f) The Fourier spectrum is composed of the frequencies $\omega_k =2k+1$, with $k\geq0$, and is time dependent. Two snapshots are analysed around the times specified in plot (d) with vertical dashed and dotted lines. The Fourier content interpolates in time between exponentially decaying spectra dominated by either $\omega = 3$ or $\omega = 5$, as shown in the bottom plots (e) and (f).
}
\label{xChirp3and5Plot}
\end{figure}

\section{Driving at more than one resonance}
\label{morefreq}
The efficiency of the adiabatic method in building non-linear oscillons naturally raises the question about what would be the result if we apply the same procedure with a sum of two or even more resonant drivings.

To start with, we quench adiabatically from AdS using two resonant drivings. As in the case of a single frequency, the idea here is to {\em inject} the drivings, see whether a stable quasi-periodic solution is established, and then {\em extract} the drivings out. There are different protocols that can be used. The one which produces the cleanest answer consists of first producing an $\omega_\mathrm{b}=5$ oscillon ({\em i.e.} injecting and extracting such a resonant driving), and then, on it, switching on and off a periodic driving with $\omega_\mathrm{b}=3$ adiabatically. 

\begin{figure}[hb!]
\begin{center}
\includegraphics[scale=0.5, trim= 0 0.6cm 0 0.35cm, clip]{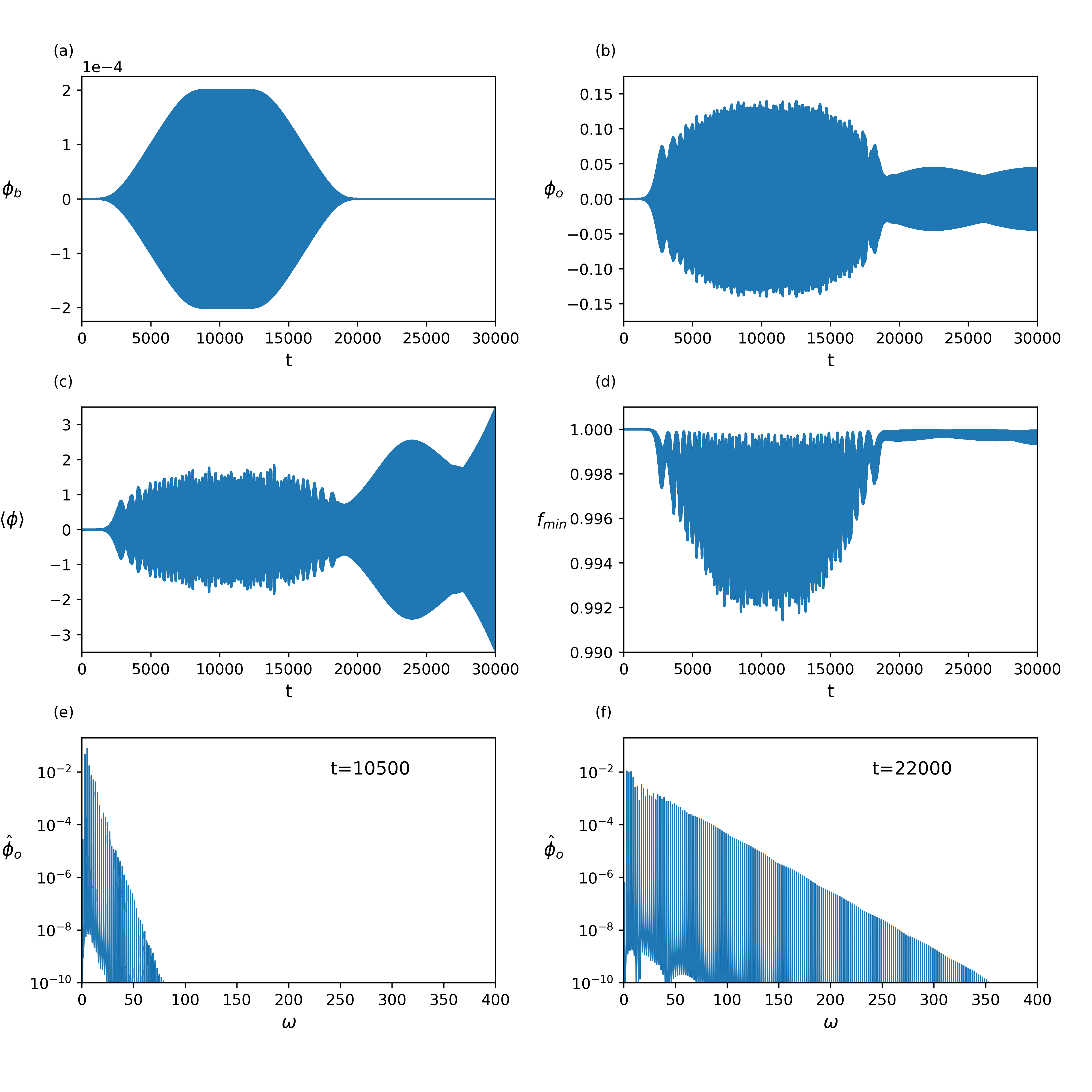}
\end{center}
\caption{Injection of simultaneous $\omega_\mathrm{b}=3$ and $5$ drivings (a). Using the $\vev$ $\langle \phi\rangle$ (c) as a diagnostic, we observe a strong interference region while the driving is on, followed by a monotonous increase after extraction. The Fourier content, shown in (e) and (f) and composed of the frequencies $\omega_k =2k+1$ with $k\geq0$, reveals that while the driving is on, the cascading towards higher frequencies is blocked. In the free evolution after extraction, the simulation seems to be consistent with non-linear turbulent behaviour leading to future collapse. However, the convergence drops monotonically from its value of 4, and the simulation stops being reliable.
}
\label{x3and5simult}
\end{figure}

The naive expectation of ending up with a non-linear two-frequency mode is wrong.\footnote{Non-linear two-mode oscillons constructed in \cite{Choptuik2018,Choptuik2019} have frequencies not related to those of linearised normal modes.} After extracting the driving, the remnant solution and the results have been collected in Fig. \ref{xQuench3and5Plot}. In plot (a) the two successive drivings at $\omega_\mathrm{b}=5$ and $\omega_\mathrm{b}=3$ are shown. Plots (b), (c) and (d) show that the final geometry is quasi-periodic in time. Plot (f) shows the time Fourier transform of the remnant solution. It is a sequence of peaks at frequencies $\omega_k = 2k+1=1,3,5,...$\footnote{To be more precise, there is an overall shift, and the correct spectrum is $\omega_k = \omega_0(2k+1)$. However, $\omega_0 = 0.9994$ is very close to 1. A similar observation affects the spectra shown in Figs. \ref{xChirp3and5Plot}(e-f) and \ref{x3and5simult}(e-f).} The amplitudes fall with an almost exponential law, starting from the dominant one, which in this case is $\omega = 3$. Our results connect nicely to the quasi-periodic solution proposed in \cite{Buchel2014hts} as a solution to the effective two-time formalism (see Figs. 1 and 2 in that reference), which also has an exponentially decaying spectrum. 

Actually, we can modify our protocol. For this, with the same driving amplitude $\rho_\mathrm{b} \sim 10^{-4}$ as before, we introduce a frequency chirp that goes all the way from $\omega = 2.8$ up to $\omega = 5.2$, thereby piercing both resonant oscillons lines with $\omega \sim 3$ and $5$ in Fig. \ref{fig:level_plot}. After extracting the driving, we look at the remnant oscillations, which are plotted in Figs. \ref{xChirp3and5Plot}(c-d). We also find here a pattern of strongly modulated quasi-periodic oscillations. The Fourier transform exhibits a time-dependent spectrum that interpolates back and forth between two quasi-periodic solutions whose spectrum is either peaked around $\omega = 3$ or $\omega = 5$ (compare again with the similar solutions to the two-time effective formalism in Fig. 1 of \cite{Buchel2014hts}).\footnote{Although not as strong and without changing the dominant frequency, this back-and-forth movement in the spectrum, shown in Figs. \ref{xChirp3and5Plot}(e-f), can also be observed in the case presented in Fig. \ref{xQuench3and5Plot} with a non-simultaneous injection of both driving frequencies.}

\subsection{Blocking the cascade}

A third protocol shows a remarkable property: we inject and extract a balanced combination of $\omega_\mathrm{b} = 3, 5$ drivings simultaneously, as shown in Fig. \ref{x3and5simult}(a). The response in the interior, in plots (b-d), shows a pattern of strong random interference while the driving is on. After the extraction of the driving, the envelope becomes smooth, but the overall oscillation amplitude starts increasing. The expectation is that the dynamics should end up collapsing into a black hole after some time. This is supported by the Fourier analysis shown in plots (e) (for the driven period) and (f) (for the free evolving period). Actually, the transition from (e) to (f) is very fast, indicating that the weak turbulent cascade sets in once the driving is switched off. In contrast, in (e), the cascade is blocked by the driving, and the spectrum drops much faster.

We have indeed checked that the regularity during the driven phase and the strong fall-down of the spectrum are features preserved for very long-time drivings. Indeed, the open periodically-driven system is stable for as long as we have been able to simulate, with fourth-order convergence. Using the boundary language for Floquet systems, our results indicate that for a sufficiently low amplitude of the resonant combined driving (here ${\cal O}(10^{-4})$), the system achieves a non-heating phase where coherence is preserved and thermalisation is avoided, much as it happened with the single-resonant-driven TPSs. From the bulk perspective, the exchange of energy-momentum through the boundary acts by blocking the turbulent cascade. It would be interesting to clarify this issue by resorting to a resonant analysis of the non-normalisable solutions, probably along the lines of \cite{Cownden2020}.

As for the final fate of the free evolution, the plots indicate a tendency towards collapse. This can be appreciated in plot (c), where the amplitude of the oscillations in $\langle \phi\rangle$ grows unbounded. Consistently, the Fourier spectrum (f) gets rapidly enriched with higher and higher modes. In accordance with this cascading, the numerical study of this free evolution is demanding. We have provided convergence tests in Appendix \ref{app: numerics} (see Fig. \ref{xconvergencias}(g)). We see that while the driving is on, a convergence factor of $4$ is achieved. However, the late cascade is not fast enough to produce a collapse into a black hole before the convergence is lost. Hence, at this point, the final fate of the 2-mode injected solution is only speculative.\footnote{In \cite{dimitrak2018}, using two-time formalism, 2-mode data were conjectured to initialise solutions collapsing in infinite time.}

\section{Driving with noise}
\label{multimode}

In the final part of the paper, we will study the response of the system to a multi-mode driving. Starting from the AdS vacuum, we will adiabatically inject the following driving on the boundary:

\be
\phi_\mathrm{b}(t) = \frac{\epsilon}{N(T)} \sum_{i=1}^{n} \phi_{n} \cos( \omega_{\mathrm{b},n} t + \varphi_n) \ , \label{phibnoise}
\ee

where $\epsilon$ is a constant, and $N(T)$ is a normalisation factor that depends on $T$. The relative amplitudes follow the distribution 
\be
\phi_{n}^2 = \frac{n_\mathrm{BE}}{\omega_n} \label{varphisn} \ ,
\ee
where $n_\mathrm{BE}=(e^{\omega_n/T}-1)^{-1}$ is the Bose-Einstein distribution, and $T$, as a sort of temperature, controls the exponential decay of the amplitudes with the frequency. The distribution in \eqref{varphisn} has been chosen such that the relative amplitudes of the time derivative of the driving, $\Pi_n=\omega_n \phi_n$, follow a Johnson-Nyquist noise distribution, 
\be
\Pi_n^2 = \omega_n n_\mathrm{BE} \ .
\ee

The power spectral density of $\Pi_\mathrm{b}$ is
\be
P(\omega) = \frac{\epsilon^2}{N(T)^2} \sum_{n=1}^\infty \Pi_n^2 \delta (\omega-\omega_n) \ ,
\ee
and the normalisation factor in \eqref{phibnoise} is set to 
\be
N(T)^2 = \sum_{n=1}^\infty \Pi_n^2 \ .
\ee
This choice is useful in that the total power is now independent of $T$,
\be
P = \int_{-\infty}^{\infty} P(\omega) d\omega = \epsilon^2 \label{totpow} \ .
\ee

On the other hand, relative phases are random and follow a uniform distribution $\varphi_n \in [0,2\pi)$. This driving corresponds to a kind of thermal noise composed only of resonant modes. 

Our original motivation when we started the present study was to model in this way a holographic open system coupled to a statistical ensemble bath. In our setup, $T$ would play the role of the environment temperature. We have computed numerically several evolutions where the noisy driving has been injected from zero up to $\epsilon = 2\times10^{-4}$ for different temperatures.

\begin{figure}[ht!]
\begin{center}
\includegraphics[scale=0.6]{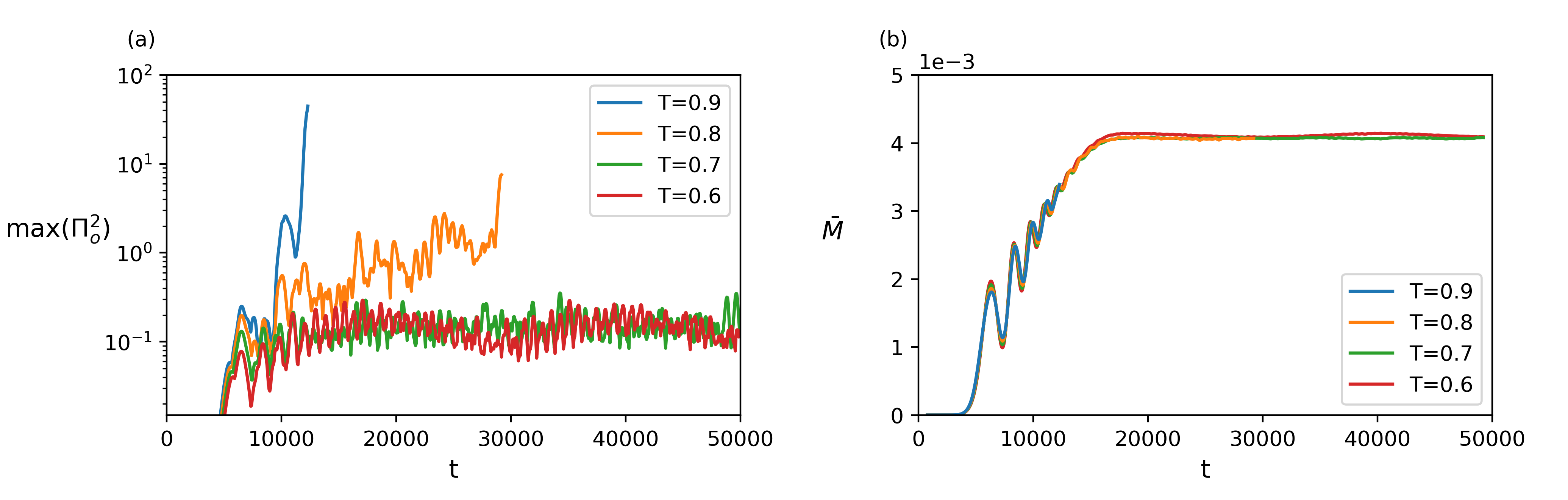}
\end{center}
\caption{ 
Multi-resonant mode driving. (a) We plot the envelope of the oscillations in $\Pi(t,x=0)^2$, which is a positive quantity. Curves stop at the point where convergence is lost in each case. (b) We show the mass $\bar M(t)$ averaged over many oscillations.
}
\label{xPio2}
\end{figure}

In Fig. \ref{xPio2} we present the results obtained for different values of the noise temperature $T$. In the left plot, max$[\Pi(t,x=0)^2]$ represents the envelope of the oscillations of the scalar curvature at the origin, $R=-2\Pi(t,x=0)/l^2-12/l^2$. Its growth was proposed in \cite{Bizon:2011gg} as a good figure of merit to diagnose the dynamics entering a collapsing regime. In the four plotted curves, the injection time is very large, $\beta=2\times10^{4}$, in order to avoid overheating due to the injection quench. There appears to be a threshold value of T $\sim$ 0.7 above which the dynamics absorbs energy monotonically right from the very start. Presumably, the end result should be the collapse and formation of a black hole. Unfortunately, the convergence is lost before we can reach any firm conclusion. We stopped the simulation whenever the convergence factor dropped below 2. 

In contrast, below that threshold, the regular dynamics suddenly extends for as long as we have been able to simulate. Fourth-order convergence is maintained all the time. Notice that below such temperature, the amplitude of the oscillations in $\Pi(t,x=0)$ saturates at a certain value well before the end of the injection. This does not rule out a later regime where the dynamics destabilises and starts drifting slowly upwards. For the sake of clarity, the number of active modes (i.e. those which play a role above the machine precision value) ranges from 22 at $T=0.6$ to 34 for $T=0.9$.

The mass does not seem to be such an accurate diagnostic of the route to collapse. In the right plot in \eqref{xPio2}, we represent the mass averaged over many periods of oscillation to avoid cluttering. The four temperatures show very similar behaviour, only distinguished among themselves by the stopping time, controlled by the convergence factor.

\section{Conclusions}

In this paper, we have studied the response of a simple holographic system to a periodic change in the boundary conditions. This system involves a plain real scalar field backreacting and a 4-dimensional gravity with a negative cosmological constant. The dual of this setup is a strongly coupled CFT subject to a periodic shaking of a coupling constant, a.k.a. Floquet system. We have extended and completed aspects of the analysis that was initiated in (the second half of) \cite{biasi2018floquet}. There, the existence of {\em time-periodic solutions} (TPSs) was put on firm grounds by constructing them in two different ways: explicit solution via pseudo-spectral methods and adiabatic preparation from the AdS vacuum. Our focus in this paper has been set on the second method. TPSs exist for generic values of the amplitude and the frequency of the driving. They are neither dual to equilibrium states nor non-equilibrium steady states (NESS) in the dual QFT. In fact, their mass is oscillating in time. Yet they represent the preferred states compatible with the given boundary conditions, and their mass averaged over a period is indeed the lowest. In general, this average mass is positive, namely, above the AdS vacuum energy. Curiously, for drivings of low enough frequency, $\omega \leq 1.25$, the average mass is negative while the periodic driving is on. 

Given these premises, an effect that very well deserves the name of `Floquet adiabatic theorem' was anticipated in \cite{biasi2018floquet}. It stated that one can move continuously along a path in the space of TPSs as long as the characteristic time of the modulation in amplitude and frequency is long in comparison with the gap of the lowest normal mode fluctuation of the TPSs along the path. There are vast regions where this occurs (see grey areas of linear stability in Fig. 33 in \cite{biasi2018floquet}). In particular, this precludes the modulation that crosses oscillon lines. These are lines where the scalar-gravity system oscillates without the need for any driving. They have a zero mode (the deformation mode along the line), and hence, near these solutions, we expect a non-linear resonant generation of higher harmonics. This is in essence the core of the present study.

The prototypical modulating experiment we have performed in this paper involves the initial adiabatic {\em injection} of a periodic driving and the final adiabatic {\em extraction} down to zero driving. In between, several types of modulations have been studied. For example, a slow change in frequency either in the upward or downward direction. If the chirp does not traverse any oscillon line, the system returns smoothly down to the vacuum state AdS.  However, if the modulation moves the frequency crossing an oscillon line, after extraction of the driving, an oscillon of the crossed branch remains with a given mass that depends on the details of the modulation. This makes this protocol a powerful method of fabricating oscillon solutions. There is a variety of possibilities depending on both the traversing speed and the sense (either upwards or downwards). Actually, the downward frequency modulation is so efficient that, in principle, is able to generate up to the maximum stable oscillon solution and beyond, hence leading to a black hole formation dual to the loss of coherence and thermalisation in the dual QFT.

As a valuable figure of merit, we have carefully looked at the Fourier spectral analysis of the value of the scalar field at the origin $\phi(t,0)$. This is essential, in particular, to establish the purity of the oscillons that are left behind after the extraction of the resonant driving. It also has been an essential tool in exploring the remnant state left behind after the injection and extraction of a two-mode driving. Indeed, the driving with frequencies $\omega = 3$ and $5$ has been introduced in several different ways (one after the other or simultaneously) with different long-time results. The resulting solution is in general quasi-periodic, and the spectrum shows the same exponential damping found for regular solutions to the free non-linear system in the two-time formalism \cite{Buchel2014hts}. The dominant frequency can be $\omega = 3$, $\omega=5$ and even alternate between both (Fig. \ref{xChirp3and5Plot}).

An interesting feature revealed in the course of the study is the fact that the Fourier spectral decay is much steeper in the open system than in the closed system case. In a sense, the periodic driving acts by filtering out the non-linear generation of higher harmonics. As soon as it is turned off, the weak turbulent cascading sets in. The long-time fate afterwards is difficult to foresee as the numerical simulations lose convergence before any conclusive result is reached. It would be interesting to prove this effect more explicitly, maybe along the lines pursued in \cite{Cownden2020}.

Finally, a natural extension of the above analysis is the multi-resonant driving. The spectral content that we have introduced corresponds to coupling the system to external thermal noise characterised by some temperature. The overall amplitude has been set very small, and yet, we see a noticeable effect of the temperature on the long-time behaviour of the system. 

\subsection*{Acknowledgements}
We would like to thank Anxo Biasi and Alexandre Serantes for useful exchange and conversations.

This work was supported by Xunta de Galicia (Centro Singular de Investigacion de Galicia accreditation 2019-2022 and grant ED431C-2021/14),  the Mar\'ia de Maeztu Units of Excellence program under project CEX2020-001035-M,
the Spanish Research State Agency under grant PID2020-114157GB-I00, and by the European Union FEDER. The work of DTM has been supported by the Xunta de Galicia action `Axudas de apoio \'a etapa predoutoral' under the grant ED481A-2020/106.

\appendix
\section{Details on the numerics}
\label{app: numerics}

In this appendix, we summarise the basic aspects of our numerical simulations. The equations of motion follow from the action \eqref{com_act} with the ansatz \eqref{line1}. The effective system of equations is conveniently expressed in terms of the following variables: 

\be
\Phi \equiv \phi' \ , ~~~~~~~
\Pi \equiv \frac{e^{\delta}}{f} \dot\phi \ , 
\ee

where $\dot{}$ and $'$ represent the time and spatial derivatives respectively. We are interested in time-periodic solutions with harmonic boundary conditions such that $\phi(t,\pi/2) = \rho_\mathrm{b} \cos(\omega_\mathrm{b} t)$. 
We work in the boundary gauge, $\delta(t,\pi/2)=0$. We also have that $\Phi(t,\pi/2)=0$ due to the asymptotic near-boundary expansion. 
With these definitions, the equations of motion become 
\begin{eqnarray}
\dot{\Phi} &=& (f e^{-\delta} \Pi)' \ , \label{eqPhireal}\\
\dot{\Pi} &=& \frac{1}{\tan^{2} x}(\tan^{2} x f e^{-\delta} \Phi)' \ , \label{eqPireal}\\ 
\delta' &=&-\displaystyle \sin x \cos x \left( \Phi^2 + \Pi ^2 \right) \ , \label{edeltareal}\\
f ' &=& \frac{1+2\sin^{2} x}{\sin x \cos x} ( 1 - f) + f \delta' \ , \label{eqFreal}
\end{eqnarray}

and the Momentum Constraint is 
\be
\dot{f} + 2 \sin x \cos x f^2 e^{-\delta} \Phi \Pi = 0 \ .\label{MC}
\ee

The driving protocol deserves some comments. The boundary conditions and their derivatives must be continuous functions of time. To fulfil this condition we use the following non-analytic function to modulate the amplitude of the scalar field boundary condition:
\be
q_{\pm}\left(t,\beta,\sigma\right)=\frac{1}{2}\left(1\pm\tanh\left(\sigma\,\beta\left(\frac{1}{t-t_{0}}+\frac{1}{t-\left(t_{0}+\beta\right)}\right)\right)\right) \ .
\ee
This function has the property that both at the beginning ($t=t_0$) and the end ($t=t_0+\beta$) of the quench all its derivatives are zero. This is crucial to guarantee that not only time derivatives are continuous, but also spatial derivatives are the correct ones to fulfil the corner conditions \cite{biasi2018floquet,Horowitz_Wang}. The parameter $\sigma$ controls the slope of this function. For amplitude modulations we set $\sigma=\frac{1}{2}$.

In order to implement the chirp protocol, where the frequency is shifted, we fix boundary conditions of the form
\begin{eqnarray}
\phi(t,\pi/2) &=& \rho_\mathrm{b} \cos\left(\nu\left(t\right)\right) \ , \label{chirpfun}\\
\nu\left(t\right) &=& \left[\omega_{1}+\left(\omega_{2}-\omega_{1}\right)q_{-}\left(t,\beta,\sigma=\frac{1}{4}\right)\right]t-\frac{\beta}{2}\left(\omega_{2}-\omega_{1}\right)q_{-}^{2}\left(t,\beta,\sigma=\frac{1}{4}\right) \ . \nonumber
\end{eqnarray}

With this definition, the instantaneous frequency will be $\omega(t) = d\nu(t)/dt$, an example of which can be seen in Fig. \ref{xchirpup}(b).

\begin{figure}[htp!]
\begin{center}
\includegraphics[scale=0.5, trim= 0 1.5cm 0 1.5cm, clip]{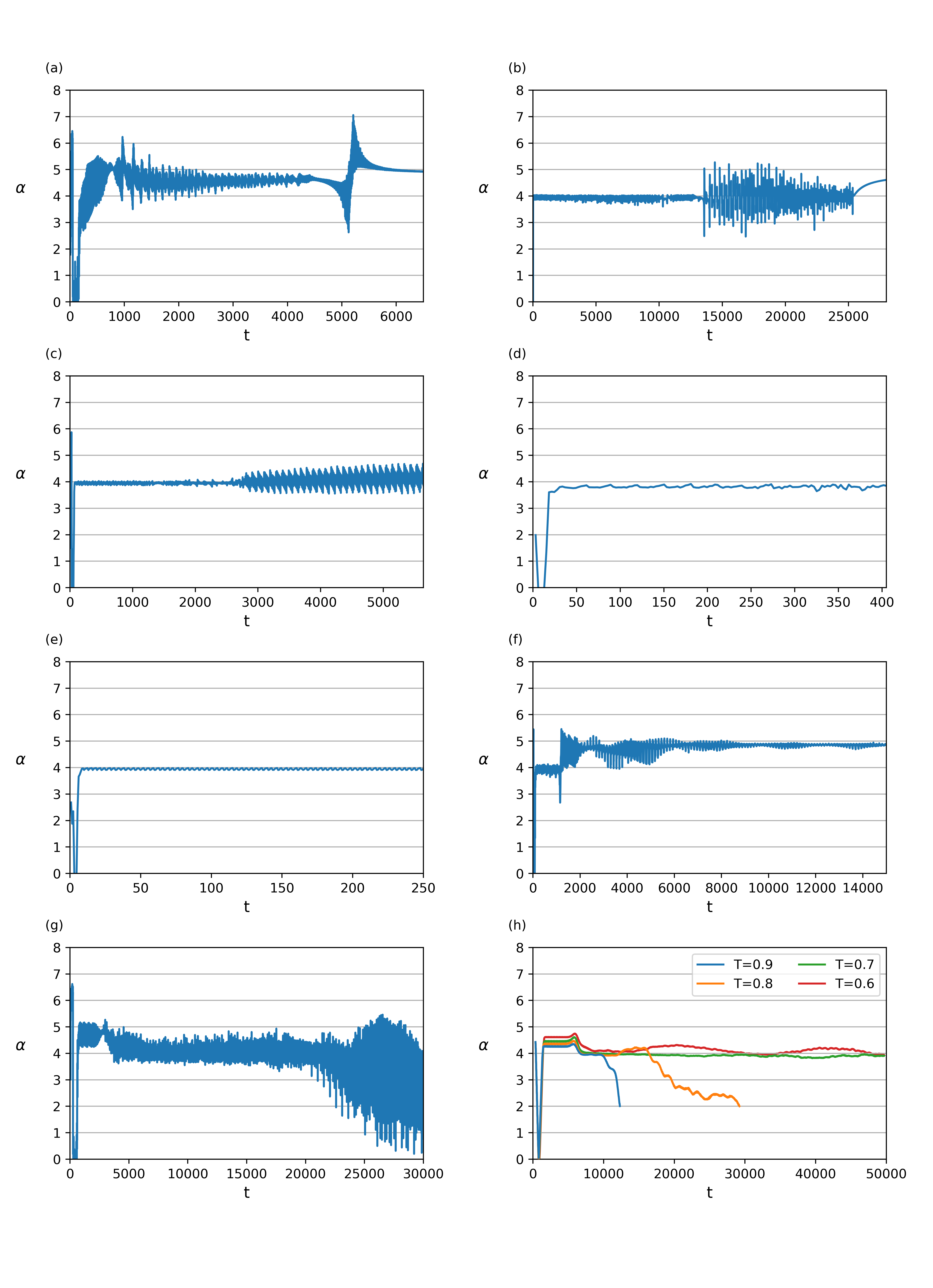}
\end{center}
\caption{Order of convergence $\alpha$. (a) Fig. \ref{xQuench3Plot}: driving at resonance $\omega_\mathrm{b}=3$, (b) Fig. \ref{xchirpdownn}: down-chirping, (c) Fig. \ref{chirp_no_line}(a): chirping without crossing any resonance curve,  (d) Fig. \ref{xzero_vev}: zero $\vev$ solutions, (e) Fig. \ref{fig:xMasa_negativa}(b): energy-extracting solutions, (f) Fig. \ref{xChirp3and5Plot}: chirping across two resonances, (g) Fig. \ref{x3and5simult}: simultaneous $\omega_\mathrm{b}=3$ and $5$ drivings, (h) Fig. \ref{xPio2}: multi-resonant mode driving.
}
\label{xconvergencias}
\end{figure}

The discretisations have been performed at order 4 in accuracy. For time marching, we have used RK4 and a spatial resolution of $2^{12}$ points. To check for convergence, we evolved the same initial data and boundary conditions using three different grid spacings. Being $g_{a}\left(t,x\right)$ a certain quantity obtained from the evolution on the $\frac{\pi/2}{a}$ grid, we can compute the norm
\be
\left\Vert g_{a}-g_{b}\right\Vert _{c} \equiv \left(\int_{0}^{\pi/2}dx\left(g_{a}-g_{b}\right)^{2}\right)^{1/2} \ ,
\ee
where $\left\Vert \right\Vert _{c}$ denotes that the integration is performed on the grid with $\frac{\pi/2}{c}$ spacing. We measure the convergence with the rate
\be
Q = 2^{-\alpha} = \frac{\left\Vert g_{4n}-g_{2n}\right\Vert _{n}}{\left\Vert g_{2n}-g_{n}\right\Vert _{n}} \ ,
\ee
where $\alpha$ is the order of convergence. 
We compute both the numerator and the denominator by adding the contributions of $\Pi$ and $\Phi$. 

It is natural to expect a convergence factor close to $4$, although we could tolerate a somewhat smaller value. We have performed convergence tests in all the numerical experiments reported, and the results are shown in Fig. \ref{xconvergencias}. After an initial transient, in general, the convergence factor indeed approaches the value 4, and even surpasses it in certain cases. This is a very satisfactory signal of confidence in the numerical validity of our results. We have even used this criterion to stop the simulation, especially when approaching a collapse.

\bibliographystyle{JHEP}

\bibliography{library.bib} 

\end{document}